\documentclass{aastex63}

\usepackage{color}

\def \tcr{\textcolor{red}}

\def\beq{\begin{equation}}
\def\eeq{\end{equation}}
\def\c{{\cdot}}
\def\t{\ {\theta}}
\def\b{\ {\beta}}
\def\D{\ {\Delta}}
\def\g{\ {\gamma}}

\shorttitle{Cosmological evolution of Blazars}
\shortauthors{Zeng, Petrosian and Yi.}

\begin{document}

\title{Cosmological Evolution of Fermi-LAT Gamma-ray Blazars Using Novel Nonparametric Methods}

\correspondingauthor{Vah\'{e} Petrosian}
\email{vahep@stanford.edu}

\author{Houdun Zeng}
\affiliation{Key Laboratory of Dark Matter and Space Astronomy, Purple Mountain Observatory, Chinese Academy of Sciences, Nanjing 210023, People's Republic of China}

\author{Vah\'{e} Petrosian}
\affiliation{Department of Physics and Kavli Institute for Particle Astrophysics and Cosmology, Stanford University, 382 Via Pueblo Mall, Stanford, CA 94305-4060, USA}
\affiliation{Department of Applied Physics, Stanford University, 382 Via Pueblo Mall, Stanford, CA 94305-4060, USA}

\author{Tingfeng Yi}
\affiliation{Department of Physics, Yunnan Normal University, Kunming 650500, People's Republic of China}



\begin{abstract}

Multi-wavelength analyses of spectra of active galactic nuclei (AGNs) provide useful information on the physical processes in the accretion disk and jets of black holes. This, however, is limited to bright sources and may not represent the population as a whole. Another  approach is through the investigation of the cosmological evolution of the luminosity function (LF) that  shows varied evolution (luminosity and density) at different wavelengths. These differences and the correlations between luminosities
can shed light on the Jet-accretion disk connection. Most such  studies use forward fitting parametric methods that involve several functions and many parameters. We use non-parametric, non-binning methods developed by Efron and Petrosian, and Lynden-Bell, for obtaining unbiased description of the evolution of the LF, from data truncated by observational selection effects. We present analysis of the evolution of gamma-ray LF of blazars with main focus on flat spectrum radio quasars (FSRQs). This requires analysis of both gamma-ray and optical data, essential for redshift measurements, and a description of the joint LF. We use a new approach which divides the sample into two sub-samples, each with its own flux limit. We use the Fermi-LAT and GAIA observations, and present results on the gamma-ray LF and its evolution, and determine the intrinsic correlation between the gamma-ray and optical luminosities corrected for the well known false correlation induced by their similar redshift dependence and evolution of the two luminosities. We also present a direct estimation of the contribution of blazars to the spectrum of the extragalactic gamma-ray background.

\end{abstract}

\keywords{Gamma-rays(637), Blazars(164), Flat-spectrum radio quasars(2163), Cosmic background radiation(317), Cosmological evolution (336), Luminosity function (942)}


\section{Introduction}
\label{sec:intro}

Blazars are the most extreme class of active galactic nuclei (AGNs), and the largest class of
extragalactic sources detected by the Large Area Telescope (LAT) of the Fermi Gamma-Ray Observatory \cite[e.g.][]{2010ApJS..188..405A,2020ApJS..247...33A,2020arXiv200511208B}. Extensive analyses of the spectral characteristics of brighter blazars with multiwavelength coverage have confirmed that the gamma-ray emission is an essential observational tool for understanding the physics of the central engines of AGNs dominated by the radiation from the plasma in the jet.
Investigations of the {\it distribution} of their spectral (and other) characteristics, and their cosmological evolutions, based on the population as a whole, can provide important information for constraining emission models. More specifically, the study of the cosmological evolution of the luminosity function (LF) and formation rate, described by the general bi-variate function $\Psi(L,z)$, can shed light also on the growth of central black holes of galaxies \citep[e.g.][]{2007ApJ...659..958D}.
It is therefore important to carry out a detailed analysis of the cosmological evolution of Blazars.
A fundamental requirement for this task is the knowledge of observational selection effects which truncate the data and can bias the results. The simplest way of accounting for this effect is to secure a sample of sources that is statistically complete with a well defined flux limit; in this case the gamma-ray limits, $F_{{\rm lim}, \gamma}$.
However, the determination of the cosmological evolution cannot be carried out based on gamma-ray observations alone; it requires source redshifts, which can be obtained by optical spectral observations, so that the optical LF and its cosmological evolution also come into play requiring determination of the tri-variate distribution $\Psi(L_\gamma, L_{\rm opt}, z)$.
The optical observations introduce additional data truncation and bias which can be accounted for with the knowledge of the optical flux limit, $F_{\rm lim,opt}$ \citep[e.g.][]{2011ApJ...743..104S}.

Correcting for this  truncation introduced by these limits, usually referred to as the Malmquist or Eddington bias, has a long history.
Majority of works dealing with this bias for the investigation of the LF and its evolution use  parametric forward fitting methods, whereby a set of assumed {\it parametric} functional forms of the LF and its evolution are fit to the data to determine the “best fit values” of the parameters. The disadvantages of such methods is that they often involve many parameters raising
serious questions about the {\it uniqueness} of the results, and often require
binning of the data and thus large samples. Via such methods the evolution of the gamma-ray luminosity function (GLF) have been described by  the parameterized form: e.g. the forms of pure density evolution (PDE), pure luminosity evolution
(PLE), and luminosity-dependent density evolution (LDDE)
(\citep[e.g.][]{2012ApJ...751..108A,Ajello_2013,2014ApJ...780..161D,2014ApJ...786..129D,
2012ApJ...749..151Z,2013MNRAS.431..997Z,2019MNRAS.490..758Q}).
There are, however, several  {\it non-parametric, non-binning methods} (e.g.~the $V/V_{\rm
max}$, \cite{1968ApJ...151..393S}; $C^-$, \cite{1971MNRAS.155...95L}) that have been used successfully for studies of AGN.

However, as pointed out by \cite{1992scma.conf.....F}, a major drawback of these and other  non-parametric methods is the {\it ad hoc} assumption of {\it independent or uncorrelated
variables}; here $L$ and $z$, which means the assumption of no cosmological luminosity evolution. To overcome this shortcoming,  \cite{1992ApJ...399..345E} (EP)   developed a new method, whereby one first determines whether $L$ and $z$
are correlated or not. If correlated, then it introduces a new variable $L_0\equiv L/g(z)$ and finds a luminosity evolution function, $g(z)$, that yields an uncorrelated $L_0$ and $z$, whose distributions can be obtained using e.g.~the $C^-$ method.%
\footnote{\cite{2012msma.book.....F} have coined the name LBW for the $C^{-}$ method, which stands for Lynden-Bell-Woodroofe, for Woodroofe's further mathematical development of the method (1985, Annals of Stat). The Efron-Petrosian and LBW methods are implemented in R/CRAN package DTDA; (cran.r.project.org/web/package/DTDA/DTDA.pdf).}
The correlation is evaluated using the standard ranking technique with the novel feature of using source ranks among a well defined associated set of sources, instead of the whole sample, thereby correcting for the effects of truncation introduced by observations (see EP and the discussion below for more details).
{\it The advantage} of these methods (EP-L hereafter) is that they do not involve any {\it ad hoc} assumptions  and obtained the LF and all its evolutions directly from the data, non-parametrically and without restoring to binning. These nonparametric methods are being used widely and have been
proven to be  very useful  for studies of  AGNs
\cite[e.g.][]{1999ApJ...518...32M,2011ApJ...743..104S,2012ApJ...753...45S,2013ApJ...764...43S,2014ApJ...786..109S}, and gamma-ray bursts \cite[e.g.][]{2015ApJS..218...13Y,2015ApJ...806...44P,2016A&A...587A..40P,2017ApJ...850..161T}. 
We will use these methods in this work but with the a simplifying modification described below.

The procedure described above is for treatment of a bi-variate LF, $\Psi(L,z)$, but, as stressed above, for investigation of AGN evolution in a non-optical wavelengths  one needs to consider the tri-variate distribution, here $\Psi(L_\gamma, L_{\rm opt}, z)$. which involves two luminosity evolutions, $g_\gamma (z)$ and  $g_{\rm opt}(z)$, and requires a determination of the correlation between the two luminosities. As shown in
\cite[e.g.][]{2011ApJ...743..104S,2012ApJ...753...45S,2013ApJ...764...43S,2014ApJ...786..109S} the EP method can determine the correlations between all three variables.  However, here we use a more convenient and  simpler method that separates the determination  of  luminosity evolutions at the two wavelengths. Once the data is corrected for the different luminosity evolutions we then determine the $L-L$ correlation as described in \cite{2015IAUS..313..333P}. The simplification is achieved as follows: We start with the joint sample having two separate  flux limits and consisting of sources with luminosities calculated at the two wavelengths. Then for each source we calculate two maximum redshifts that each source can have and still be in the joint sample, as outlined below. The sources are then classified as {\it optically or gamma-ray limited} whichever maximum redshift is smaller. Thus, we obtain two subsamples each truncated by  its own flux limit only, and each involving a bi-variate LF. For each subsample we obtain its luminosity evolution and proceed with determinations of mono-variate distributions after accounting for the evolution. This reduces the size of the sample to nearly half of the joint sample, but still large enough for an accurate determination of all distributions.

\section{SAMPLE SELECTION}
\label{Sample}

\subsection{Gamma-Ray Data}
\label{gamma}

The Fermi-LAT collaboration \cite{2020ApJ...892..105A} (hereafter AGNLAT) presents the fourth Fermi-LAT {\bf AGN} catalog  [based on the fourth Fermi-LAT source catalog (4FGL; \cite{2020ApJS..247...33A}) (4LAC)], containing 2863 AGNs of various types
located at Galactic latitude $|b| \geq 10^o$.
Removing the 249 sources  in 4LAC (64 FSRQs, 40  BL Lacs, 136 BCUs and 9 non-blazar AGNs), which were not associated with  known AGNs, had more than one counterpart, or were flagged for other reasons  yields a clean sample of 2614 sources, 55 of which are classified as non-blazar AGNs. Thus, we end up with 2559 blazars that includes 591 FSRQs, 1027 BL Lacs and 941 BCUs.
The average energy flux threshold denoted as $S_{\rm 25,lim}$ in AGNLAT, which is related to
$\simeq 2 \times 10^{-12}$ cm$^{-2}$ s$^{-1}$, lower than the values $\simeq 3 \times 10^{-12}$ erg
cm$^{-2}$ s$^{-1}$  and $\simeq 5 \times 10^{-12}$ erg cm$^{-2}$ s$^{-1}$ in 3FGL and 2FGL, respectively (see
Fig. 15 of \cite{2020ApJS..247...33A}).
{AGNLAT, in addition to the photon flux, $F(25)$ and its power-law spectral index $\Gamma>1$  in the 100 MeV to 100 GeV energy band, give the energy flux  $S_{25}$, which are related  as}

\begin{eqnarray}
\label{eq:S-F}
F_{25}= S_{25} /E_1\times
 \left\{
\begin{array}{lcl}
 \frac{\Gamma-2}{\Gamma-1}\times  \frac{1-10^{-3(\Gamma-1)}}{1-10^{-3(\Gamma-2)}} &
& {\rm (if~~~ \Gamma \neq 2.0)} \\
\frac{(1-10^{-3})}{\ln(10^3)} &
& {\rm (if~~~\Gamma = 2.0)}\;,
\end{array}
\right.
\end{eqnarray}
where $E_{1}=1.602 \times 10^{-4}$ erg (corresponding to 100 MeV) is the lower energy limit.

\begin{figure*}
\plottwo{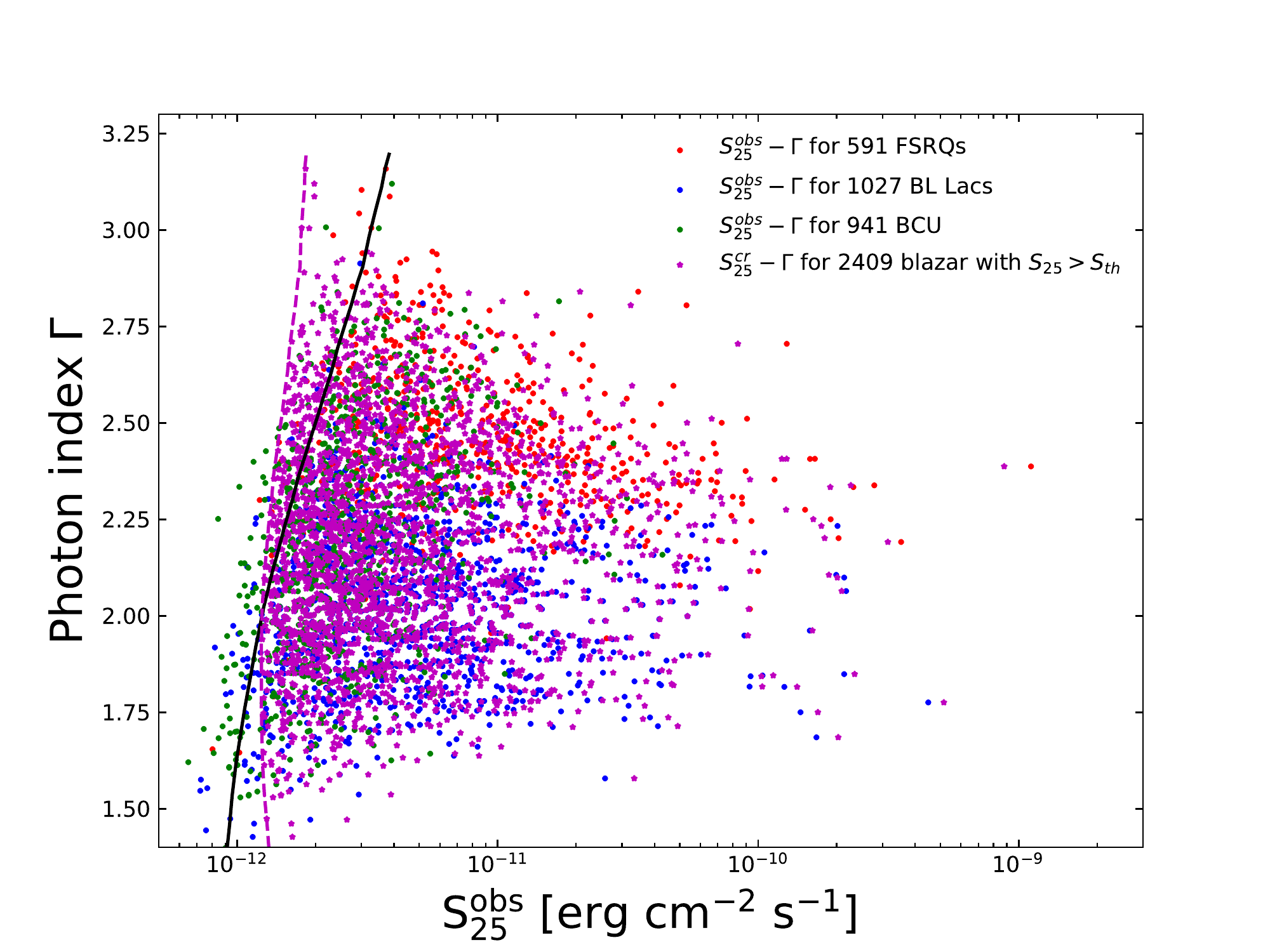}{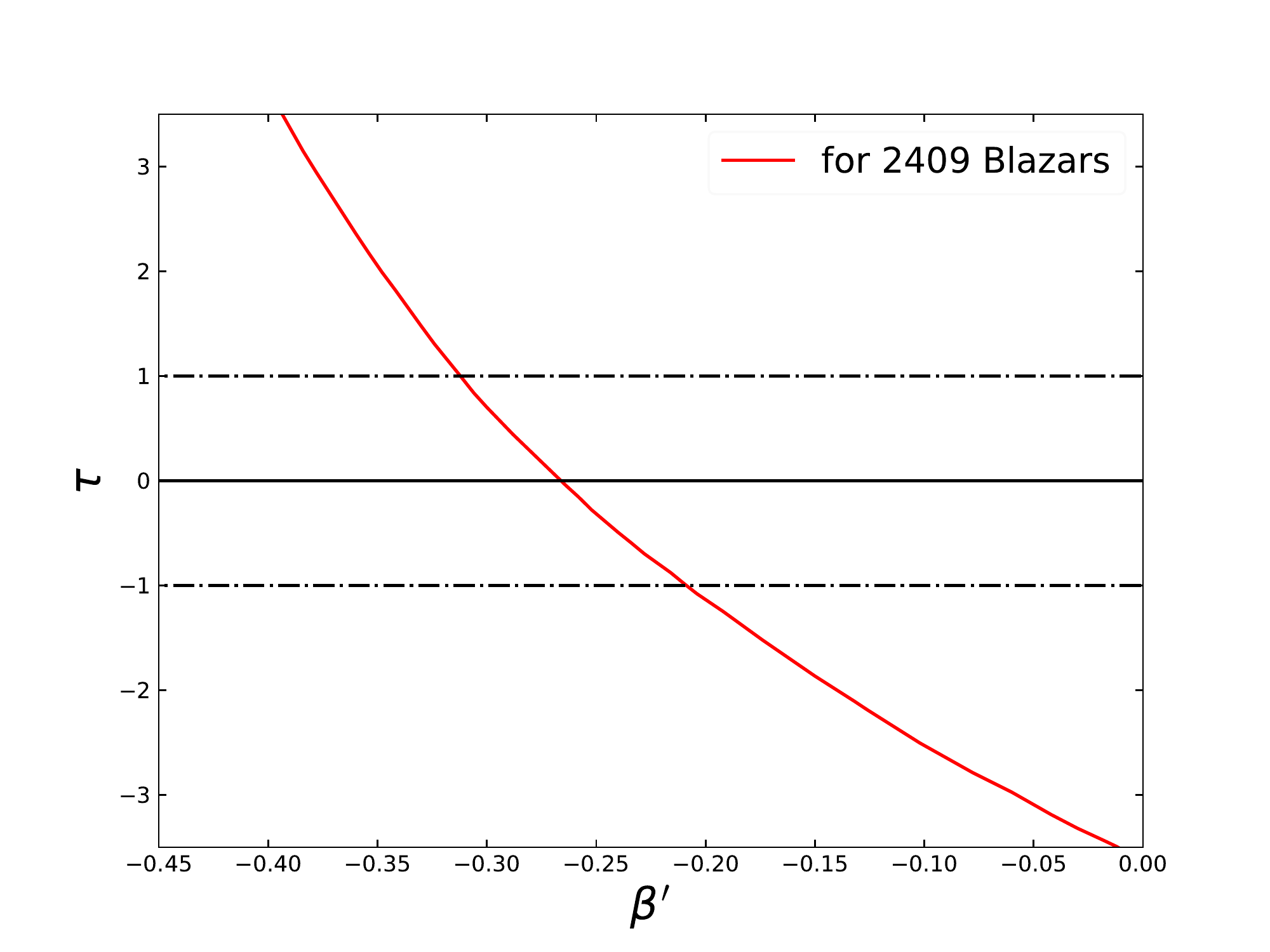}
\caption{{\bf Left:} The scatter diagram of energy flux $S_{25}$ vs power-law index  $\Gamma$.
The solid curve shows the flux threshold completeness limit $S^{\rm obs}_{25, {\rm lim}}$ vs $\Gamma$ (the truncation curve). This introduces an artificial correlation between the flux and index, which we correct for using the EP method and the Kendall $\tau$ statistics shown in the right panel. Using the best value of the correlation index $\beta^\prime$, defined in Eq. (\ref{beta'}), we calculate corrected energy fluxes and truncation curve $S^{\rm cr}_{25, {\rm lim}}$, shown by the dashed red curve. Using sources to the right of this curve we end up with corrected flux for 2409 blazars with $S^{\rm cr}_{25}> S^{\rm cr}_{\rm 25,lim}$ from  the sample of 2559 blazars.
{\bf Right:} Value of test statistic $\tau$ as a function of $\beta^\prime$ for all blazars. The values of $\beta^\prime$ for which $\tau=0$ and $\tau=\pm 1$ give the best value $-0.266$ and one sigma range $(-0.312,-0.209)$, respectively.
\label{fig:SGamma}}
\end{figure*}

It is well documented that because of strong dependence of $F_{25}$, on the {\it photon spectral index} $\Gamma$, the LAT  photon flux threshold, $F_{25,{\rm lim}}$ increases with increasing spectral index $\Gamma$, thereby introducing a bias or data truncation. Before proceeding with our goal here we need to account for this truncation. As shown by \cite{2012ApJ...753...45S} 
this bias can be accounted for using the EP method by determining the correlation between $F_{25}$ and $\Gamma$ with modified Kendall $\tau$ test statistics.
The energy flux $S_{25}$ also is affected by this bias but to a lesser degree because of the relation between $F_{25}$ and $S_{25}$ given in Eq.~\ref{eq:S-F}. Left panel of Figure \ref{fig:SGamma} shows this truncation line roughly as ${\rm Log} S^{\rm obs}_{\rm lim}=-12.0 + 0.3(\Gamma-1.5)$ (black curve) which introduces an artificial correlation between  $S^{\rm obs}_{25}$ and $\Gamma$. As done in \cite{2012ApJ...753...45S}, following the EP method, the true intrinsic correlation, corrected for this truncation can be obtained as follows: We define a corrected flux $S^{\rm cr}_{25}$ as
\begin{equation}
 {\rm Log} S^{\rm cr}_{25}= {\rm Log} S^{\rm obs}_{25} + \beta^\prime(\Gamma-2)
 \label{beta'}
\end{equation}
and evaluate modified Kendall rank statistics $\tau(\beta^\prime)$, shown on the right panel of Figure \ref{fig:SGamma}, again using the concept of associated sets. We find the best value and $1 \sigma$ range of $\beta^\prime = - 0.266^{+0.057}_{-0.046}$%
\footnote{Singal et al. (2012) used slightly different correction method. Instead of Eq. (\ref{beta'}), they used $\Gamma^{\rm cr}=\Gamma+\beta\log F_{25}$ to determine the intrinsic correlation between $F_{25}$ and $\Gamma$.}

Using the best value of $\beta^\prime$ we calculate
corrected fluxes $S^{\rm cr}_{25}$ satisfying the  truncation condition given by the dashed-red line in Figure \ref{fig:SGamma} (left), showing a very small variation of $S^{\rm cr}_{\rm lim}$ with $\Gamma$; this truncation line is essentially vertical so we adopt a constant flux limit $S^{\rm cr}_{\rm 25,lim} =1.26 \times 10^{-12}$ erg cm$^{-2}$ s$^{-1}$. This reduces the number of sources from 2559 to 2409 (568 FSRQs, 995 BL Lacs and 846 BCUs).

\begin{figure}
\gridline{\fig{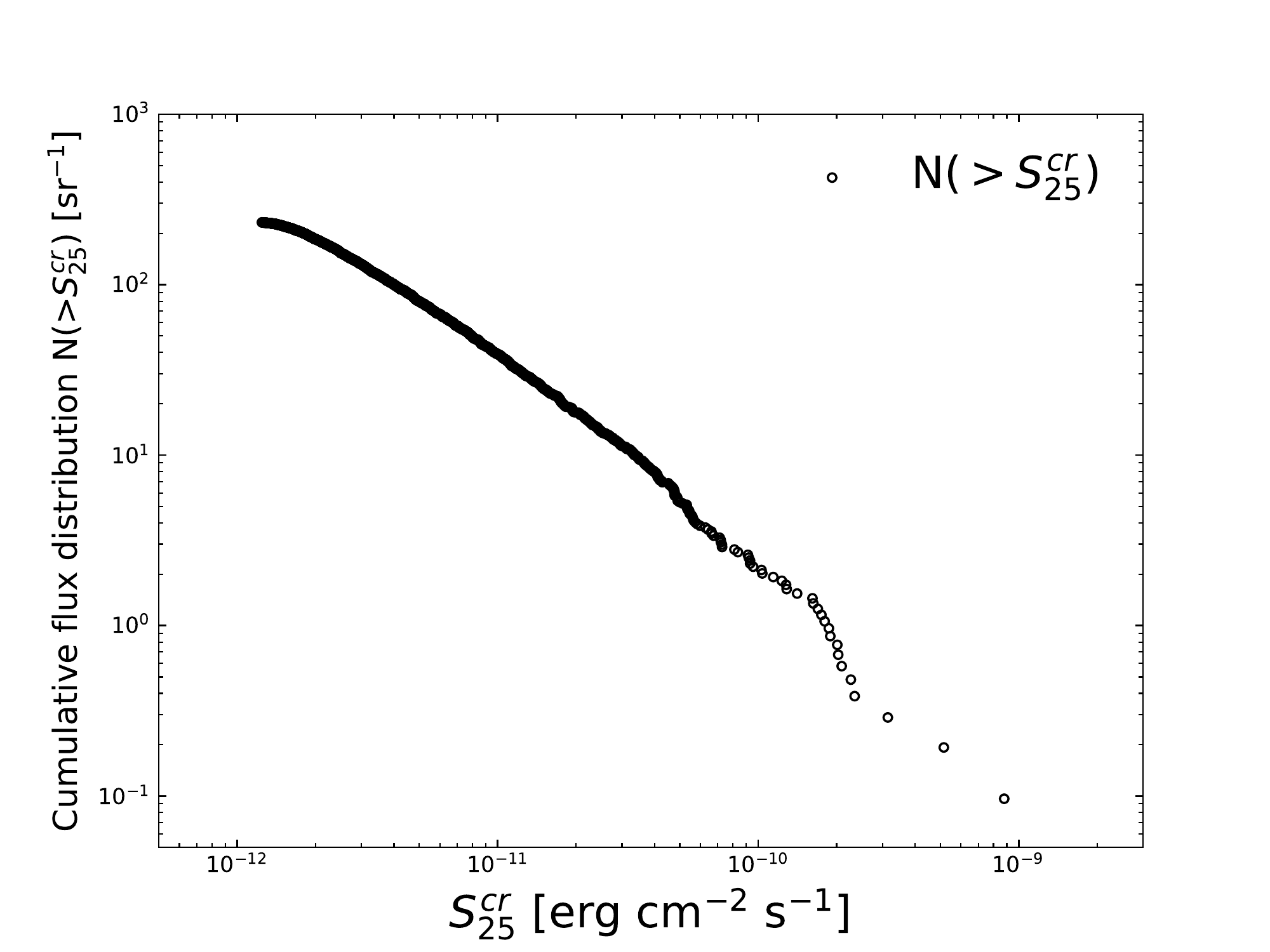}{0.32\textwidth}{}
          \fig{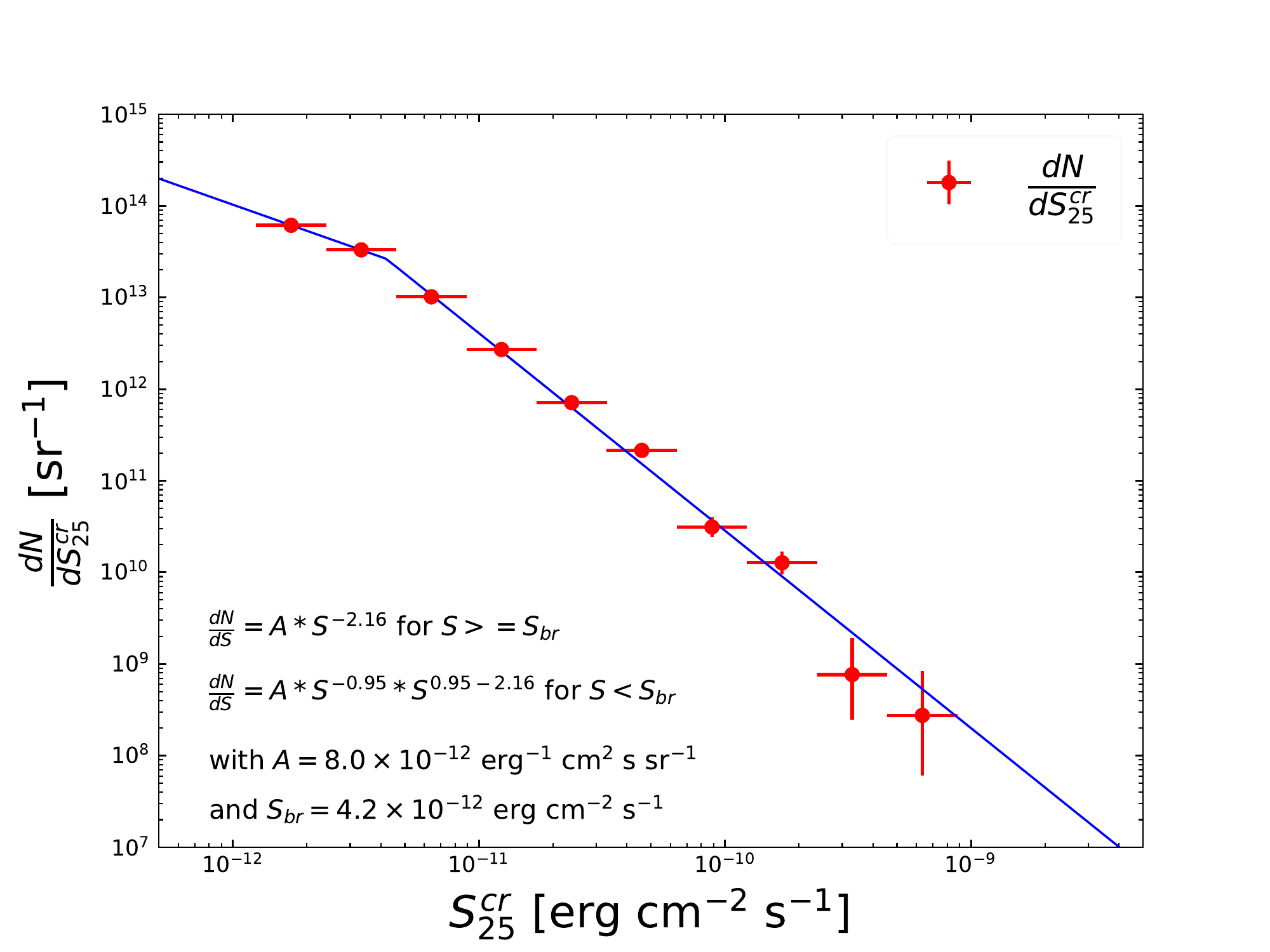}{0.32\textwidth}{}
          \fig{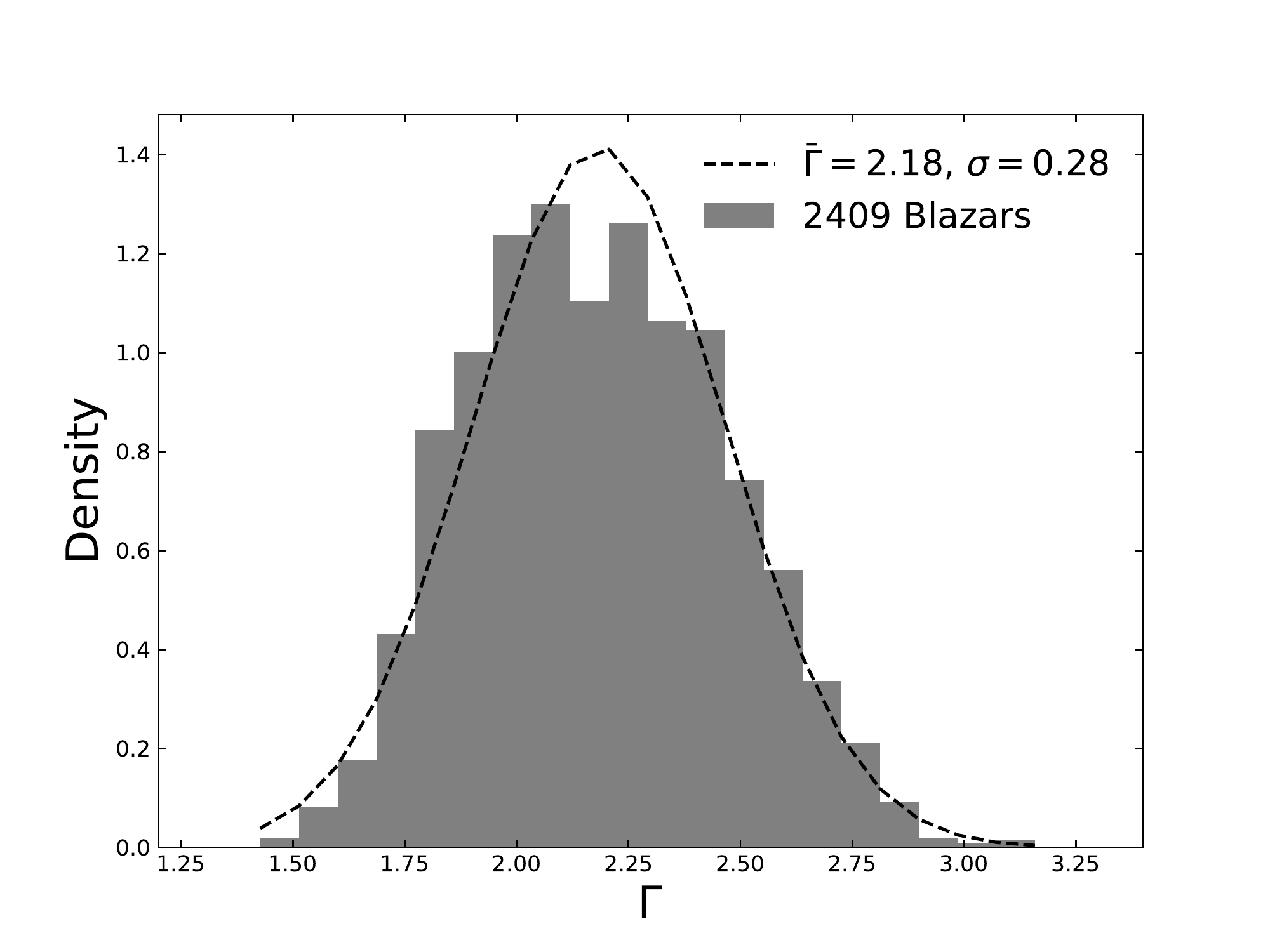}{0.32\textwidth}{}
          }
\caption{{\bf Left:} The cumulative distribution of $S^{\rm cr}_{25}$ and $N(>S_{25}^{\rm cr})=\int_{S^{\prime}}^{\infty} \frac{dN}{dS^{\prime}} dS^{\prime}$. {\bf Middle:} The differential distribution of flux $\frac{dN}{dS}$ for 2409 
Blazars which can be described by a broken power law with a break at $S_{br}=4.2 \times 10^{-12}$ erg cm$^{-2}$ s$^{-1}$. {\bf Right:} The normalized differential distribution of $\Gamma$ which can be fit well with a Gaussian of mean 2.18 and dispersion $\sigma_\Gamma=0.28$.
\label{fig:ScrGamdist}}
\end{figure}

Having established the independence of $\Gamma$ and $S^{\rm cr}_{25}$ we can now determine their mono-variate distributions. Figure \ref{fig:ScrGamdist}  shows the cumulative distribution of $N(>S^{\rm cr}_{25})$ (left panel), 
and the differential distribution $dN/dS^{\rm cr}_{25}$ (middle panel). (We will discuss the relation of these distributions to the cosmic gamma-ray background in the last section of the paper.) The right  panel of this figure shows the normalized differential distribution of $\Gamma$ which can be fit well with a Gaussian of mean $\bar{\Gamma}=2.18$ and dispersion $\sigma_\Gamma=0.28$. In what follows we will use the corrected energy fluxes and in determination of the LF and co-moving density evolution we will deal with these quantities integrated over the spectral index $\Gamma$.

Fortunately, AGNLAT also provide the \textit{Gaia} counterparts (based on a dedicated observing program), their (SEDs based) optical  fluxes and redshifts.
Out of 2409 clean sample, this yields  a sample of 1124 blazars with redshift, Gaia counterpart, $S^{\rm cr}_{25}$ and  $\Gamma$ (529 FSRQs, 501 BL Lacs and 94 BCUs) listed in Table \ref{sample}.%
\footnote{Note that 4FGL J0541.4-7334 (SSTSL2 J054212.43-7333327.8) and 4FGL J0719-4012 (1RXS J071939-401153) are not included
due to no redshift information in NASA or IPAC Extragalactic Data base (NED), and the
redshift of 4FGL J1311.0+0034 (RX J1311.1+0035) is 0.30452 \citep{2017MNRAS.464.1306C}, which can be found in NED.}
From the corrected values we calculate the gamma-ray luminosity in the rest frame 100 MeV to 100 GeV range as:
\begin{equation}
\label{Lz}
L_\gamma(z) = 4 \pi [d_{L}(z)]^2 S^{\rm cr}_{25} K=4 \pi d_{L}^2 S^{\rm obs}_{25} K^\prime,
\end{equation}
where $d_{L}(z)$ is the luminosity distance at redshift $z$ for an assumed cosmological model%
\footnote{We use the standard flat $\Lambda$CDM model with $\Omega_m=0.3$ and $H_0=70$ km s$^{-1}$ Mpc $^{-1}$.}, $K = (1 + z)^{\Gamma-2}$ is the usual K-correction term and $K^{\prime}$ is the gamma-ray corrected K-correction factor which is function of $\beta^\prime$ and $\Gamma$.
Given the value of the correlation index $\beta^\prime$, it is easy to show that the modified K-correction $K^{\prime}=(\frac{1+z}{ 1.85})^{\Gamma-2}$.
The red points in Figure \ref{fig:3} show gamma-ray luminosities vs redshift of 1096
blazars and the limiting luminosity $L_{\gamma,\rm min}(z)$ obtained from Equation \ref{Lz} with the flux $S^{\rm cr}_{25}$ replaced by limiting flux of  $S^{\rm cr}_{\rm{limit}}=1.26 \times 10^{-12}$ erg cm$^{-2}$ s$^{-1}$ (black curve),
and average spectral index $\Gamma=2.24$, yielding $K^\prime(z)=0.9\, {\rm to}\, 1.2$, for $0<z<3$, which we set to unity.

\subsection{Optical Data}
\label{optical}

We carry out similar calculation for the optical band.
To obtain luminosities we need energy fluxes. For his purpose we first obtain the magnitudes in the AB system, $G_{\rm AB}$, from the Gaia G magnitudes (wavelength range 330-1050 nm or frequencies $\nu_1=2.9\times 10^{14} {\rm Hz},\nu_2=9.1\times 10^{14} {\rm Hz}$) in the VEGMAG system as; $G_{\rm AB}=G+(G_{\rm 0,AB}-G_{0})$, where $G_{0}$'s refer to the zero point in the VEGAMAG  and
the AB systems, respectively,  given in Table 1 and 2 of \citet{2018A&A...616A...4E} for the second Gaia data release (Gaia DR2).%
\footnote{The details of these calculations can be found in the Gaia DR2 online documentation; \url{https://gea.esac.esa.int/archive/documentation/GDR2/pdf/GaiaDR2_documentation_1.2.pdf.}}
From these we obtain the  optical flux per unit frequency,  (in erg cm$^{-2}$ s$^{-1}$ Hz$^{-1}$) as, $\log f_{\nu_{\rm eff}}=-0.4(G_{AB}+48.60)$ \citep{1983ApJ...266..713O}, with $\nu_{\rm eff}= 4.81\times 10^{14}$ Hz (625 nm). Assuming a power law spectrum $f(\nu) \propto \nu^{-\alpha}$,  we calculate the "bolometric" luminosity (in erg s$^{-1}$) as
\begin{equation}
\label{Lopt}
L_{\rm opt}(z) = 4 \pi [d_{L}(z)]^2 F_{\rm opt}  (1+z)^{(\alpha_{\rm opt}-1)},
\end{equation}
where the bolometric flux is given as
\begin{equation}
\label{Fopt}
F_{\rm opt}=\int^{\nu_2}_{\nu_1} f(\nu) d\nu=\frac{(\nu_2/\nu_{\rm eff})^{1-\alpha}-(\nu_1/\nu_{\rm eff})^{1-\alpha}}{1-\alpha}\times \nu_{\rm eff}f_{\nu_{\rm eff}} \approx (1.3\pm 0.1)\times \nu_{\rm eff}f_{\nu_{\rm eff}},
\end{equation}
for spectral index $-0.5<\alpha<3$.
Gaia DR2 also provides magnitudes in two bands, $G_{\rm BP}$ and $G_{\rm RP}$, in the wavelength ranges 330-680 nm and 630-1050 nm, with the effective frequencies $\nu_{\rm BP}=5.94 \times 10^{14}$ Hz and $\nu_{\rm RP}=3.88 \times 10^{14}$ Hz, respectively. From this we can estimate the value of the optical spectral index as
\beq
\label{alphaopt}
\alpha_{\rm opt}=0.4({G_{\rm BP, AB}}-G_{\rm RP,AB})/{\rm Log}(\nu_{\rm BP}/\nu_{\rm RP}).
\eeq

\begin{figure}
\gridline{\fig{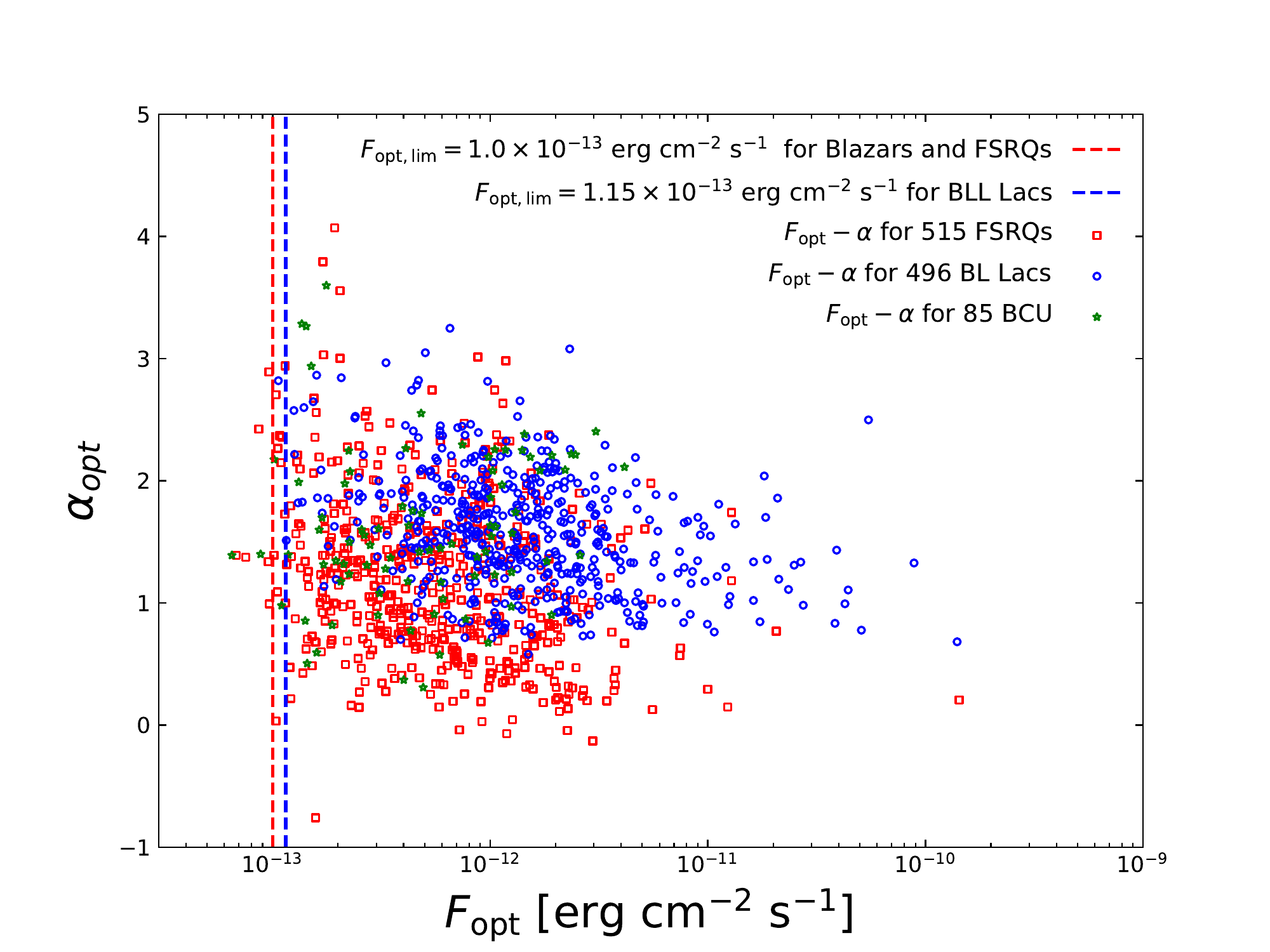}{0.32\textwidth}{}
          \fig{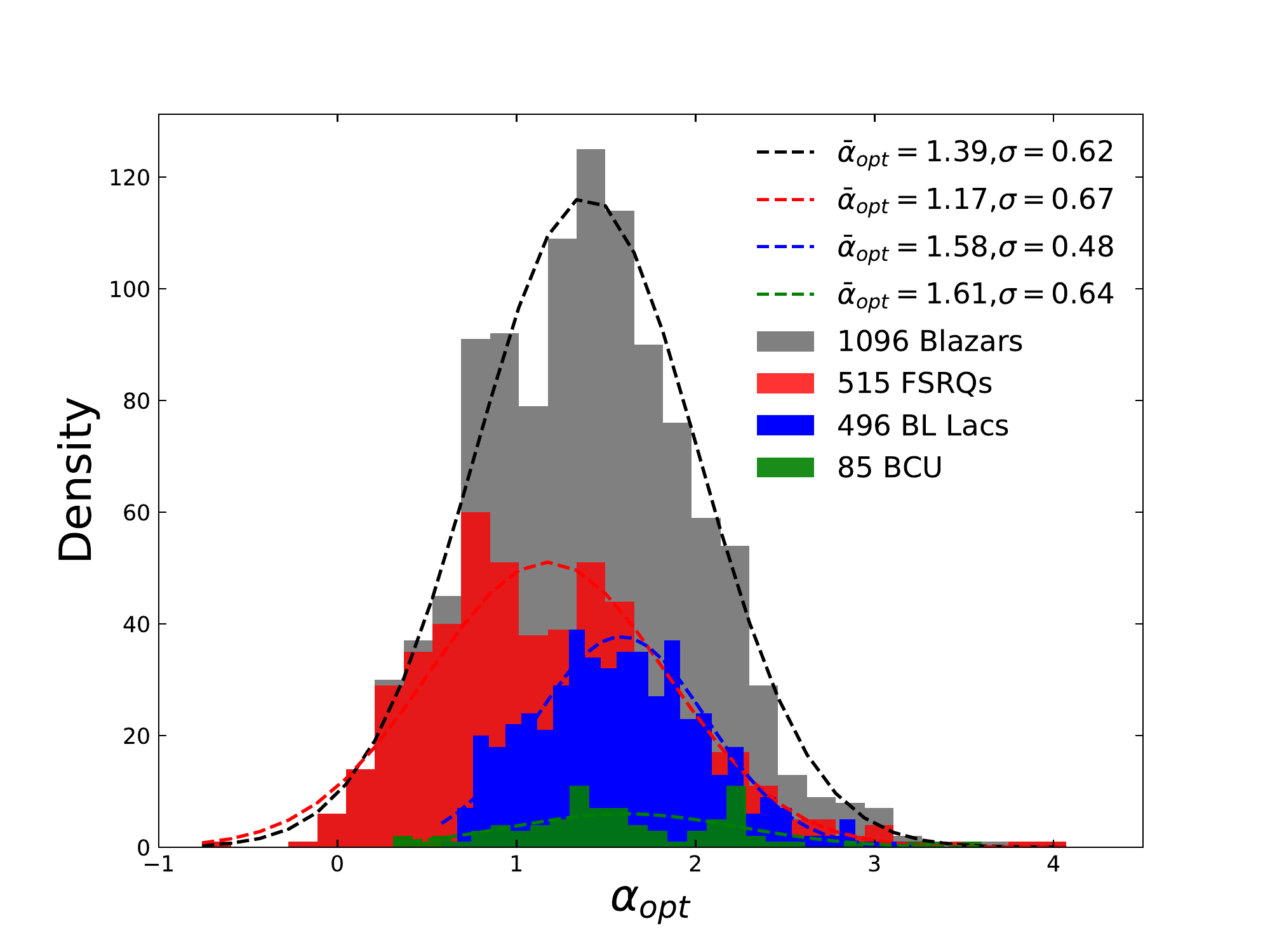}{0.32\textwidth}{}
          \fig{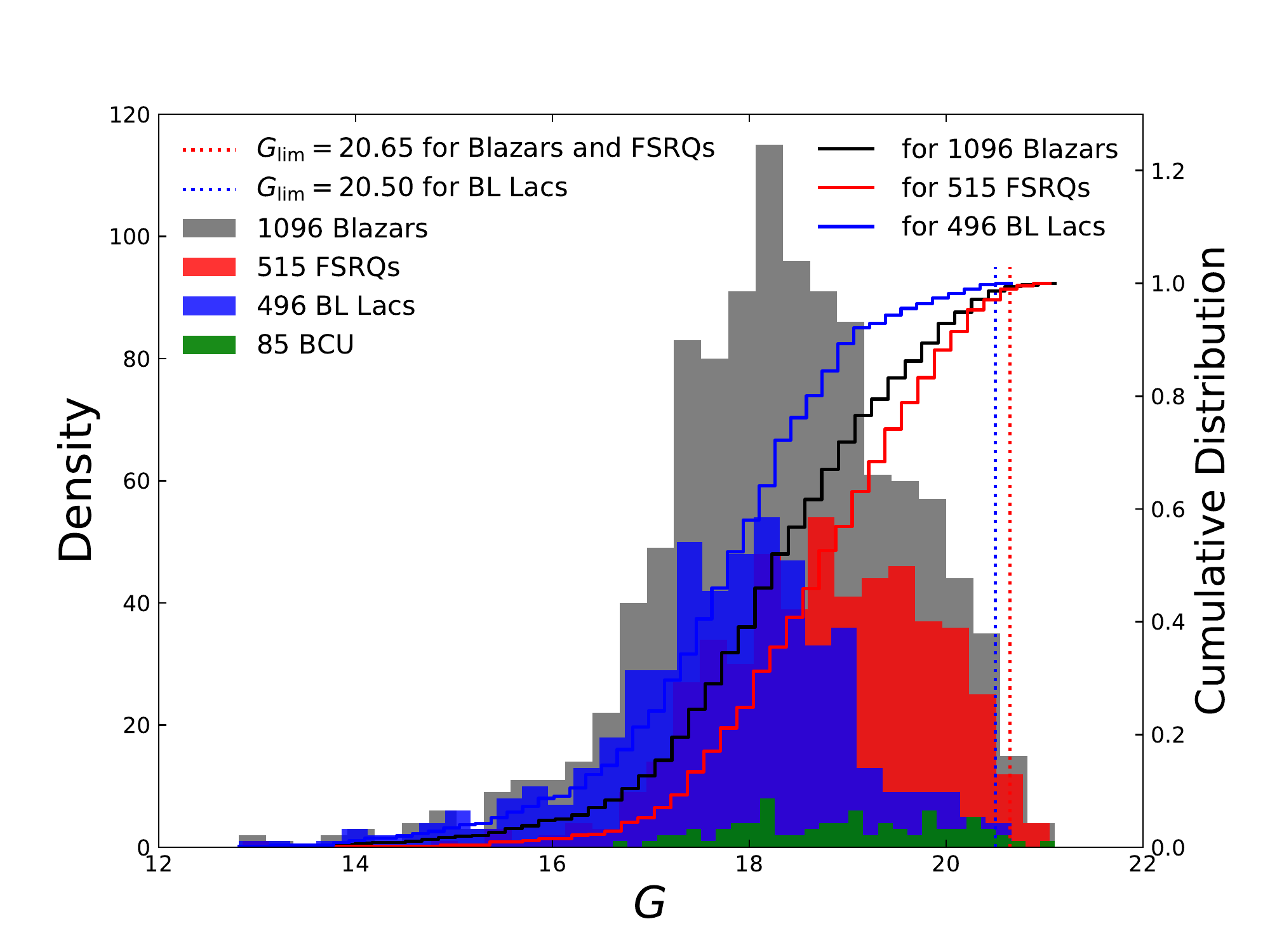}{0.32\textwidth}{}}
\caption{{\bf Left:} The scatter diagram of the optical spectra index $\alpha$ and  the "bolometric" flux, $F_{\rm opt}$, for 1096 blazars (red for FSRQs, blue for BL Lacs), indicating no (or negligible) correlation and  well defined flux limits shown by the vertical lines at $ F_{\rm opt, lim}=1.0\, {\rm and}\, 1.15 \times 10^{-13}$ erg cm$^{-2}$ s$^{-1}$, corresponding to G-magnitude limits of 20.65 and 20.50, respectively.
 {\bf Middle:} Histogram of $\alpha_{opt}$ for all blazars (in gray) showing a Gaussian distribution with the average  $\bar \alpha=1.39$ and dispersion 0.62. The red and blue histograms are for the FSRQ and BL Lac subsamples, respectively.
 {\bf Right:} The differential and normalized cumulative G-magnitude distributions of the three samples (with same color coding as above)  showing the completion limits.}
\label{fig:2}
\end{figure}

The values of these parameters are also shown in Table \ref{sample}. Note that there are 9 sources with no $G_{\rm BP}$ and $G_{\rm RP}$ in this sample. For these we use the average spectral index $\bar \alpha_{opt}=1.39$.
Figure \ref{fig:2} (left) shows the scatter diagram of the spectra index $\alpha_{\rm opt}$ and the optical flux $F_{\rm opt}$.  As evident there is no significant correlation between the flux and spectral index so that, unlike for gamma-ray fluxes, no correction is required and we can use a (conservative) constant  flux limit of  $ F_{\rm opt, lim}=1.0 \times 10^{-13}$ erg cm$^{-2}$ s$^{-1}$ (corresponding to $G_{\rm lim} \sim$ 20.65) 
for the subsample of FSRQs, and $ F_{\rm opt, lim}=1.15 \times 10^{-13}$ erg cm$^{-2}$ s$^{-1}$ (corresponding to $G_{\rm lim} \sim$ 20.50)
for the subsample of BL Lacs, shown by the red and blue vertical lines, respectively. The middle panel of this figure shows,  the distributions of $\alpha$ for 1096 blazars (gray), and subsamples of FSRQs (red) and BL Lacs (blue) with respective average values of $\bar \alpha_{\rm opt}=1.39, 1.17$ and 1.58, in good agreement with the typical optical spectral index $ 1< \alpha_{\rm opt} < 2 $ \citep{2005AJ....129.2542C}, and the results of \cite{2010AJ....139..390P}: $1.15 \pm 0.69$ and \cite{2013AJ....146..163S}: $1.17 \pm 0.30$.
The right panel of Figure \ref{fig:2} shows the differential and normalized cumulative distributions of the G-band magnitudes for the three samples with same color coding.

\begin{table}[]
    \centering
    \caption{Composition of blazars in 4LAC and Sample used in this analysis}
    \begin{tabular}{l|cccc}
          \hline
         4LAC & Blazars & FSRQs &BL Lacs & BCU \\
          \hline
         A clean sample& 2559 & 591 &1027 &941 \\
         Have redshift in clean sample &1406 &591 &657 &158 \\
         Have Gaia counterparts in clean sample &1926&553&823&550\\
         \hline
         Have redshift and Gaia counterparts in clean sample &1124&529&501&94\\
         \hline
         \hline
         A clean sample with $S_{25}>S_{25,lim}$&2409&568&995&846 \\
         Have redshift in clean sample with $S_{25}>S_{25,lim}$ &1339 &568 &666 &138 \\
         Have Gaia counterparts in clean sample with $S_{25}>S_{25,lim}$  &1850&534&811&550\\
         \hline
         Have redshift and Gaia counterparts in clean sample with $S_{25}>S_{25,lim}$ &1096&515&496&85\\
          \hline
    \end{tabular}
    \label{tab:sampleselection}
\end{table}

\begin{figure}
    \centering
    \plotone{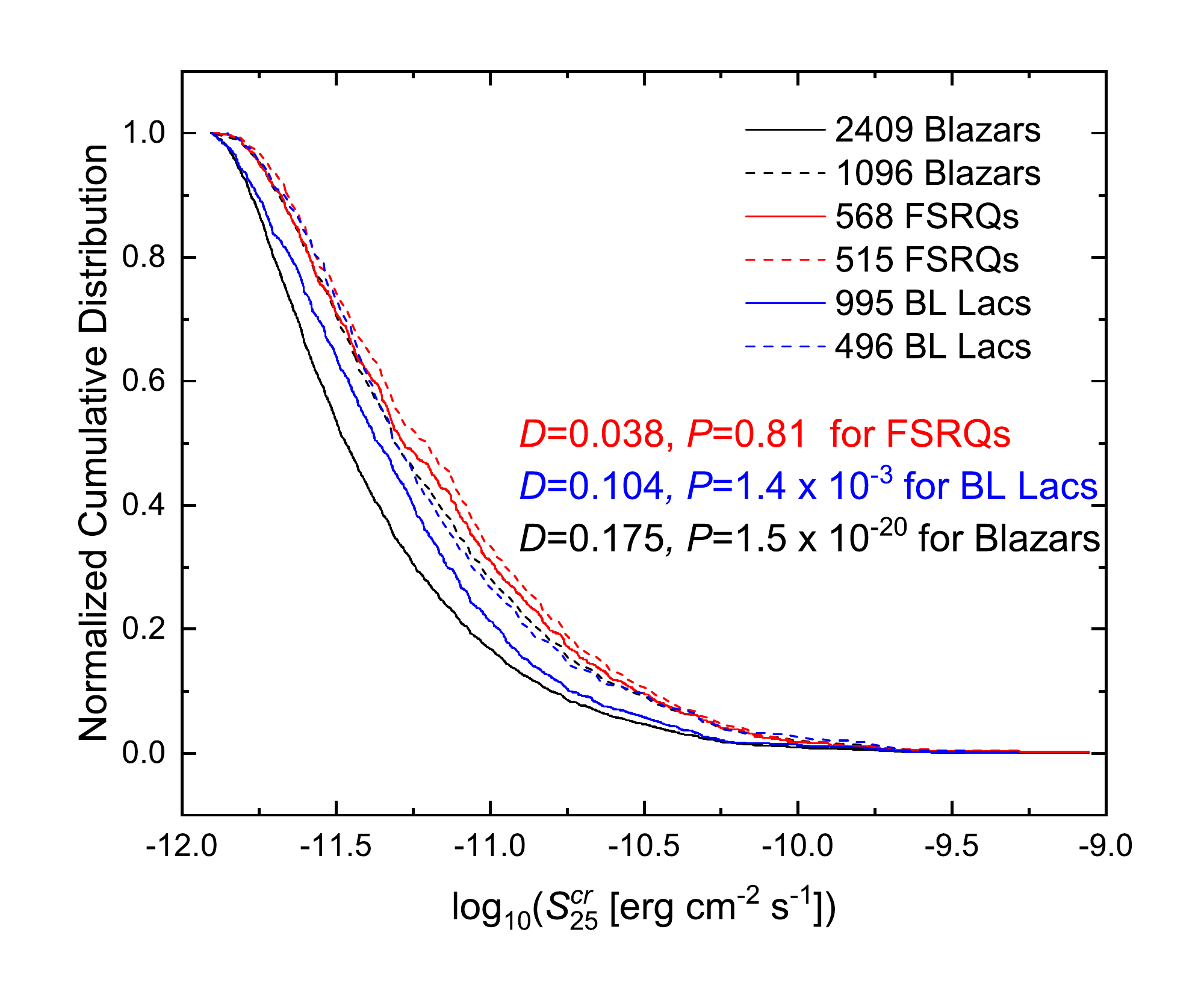}
    \caption{KS-test for two samples: sample 1 that have Gaia counterparts from a clean sample with $S_{25}>S_{25,lim}$ and sample 2 that have redshift and Gaia counterparts from a clean sample with $S_{25}>S_{25,lim}$. For blazars, the maximum deviation $D=0.175$ and confidence level $P=1.5 \times 10^{-20}$, For FSRQs, the maximum deviation $D=0.038$ and confidence level $P=0.81$  and for BL Lacs, the $D=0.104$ and $P=1.4 \times 10^{-3}$. 
    }
    \label{fig:kstest}
\end{figure}

\subsection{Combined Data}
\label{combined}

Table \ref{tab:sampleselection} shows the composition of blazars in 4LAC, various subsamples, and the samples used in our analysis. The clean sample of source that is statistically complete with well defined gamma-ray flux limit (our parent sample) is shown on row 5. The numbers in the subsample used in our investigation of the LF, that have GAIA counterparts that is statistically complete with well defined optical flux or magnitude limit, and with redshifts, are shown in the last row (our subsample). It is clear that the numbers in our sample are smaller than in parent sample. The degrees of completeness given by the ratio of the numbers in these two rows are 91\%, 50\% and 10\% for FSRQs, BL Lacs and BCUs, respectively.  With 91\% completeness the results on FSRQs in the subsample can be considered fairly robust.
To further test  this aspect, we use the Kolmogorov-Smirnov (KS) test to compare the flux distributions of the parent samples and our subsamples with redshift. Figure 4 shows the normalized cumulative distributions for the two samples for all blazars and for BL Lacs and FSRQs separately. We find maximum deviations D and p-value of 0.175 and $1.5 \times 10^{-20}$, 0.104 for all blazars and  $1.4 \times 10^{-3}$ for BL Lacs, which indicates that our subsamples for these sources are not drawn from the parent samples.%
\footnote{The large deviation for all blazars is dominated by the incompleteness of the BCUs. Omiting BCUs we find D and p-value 0.076 and $1.7 \times 10^{-3}$. We will not consider BCUs further in this paper.}
On the other hand, for FSRQs alone we find D=0.038 and p-value=0.81, a further indication that the subsamples for FSRQs with redshift is a good representatives of the parent clean sample with well defind flux limits.

\begin{figure}
\plotone{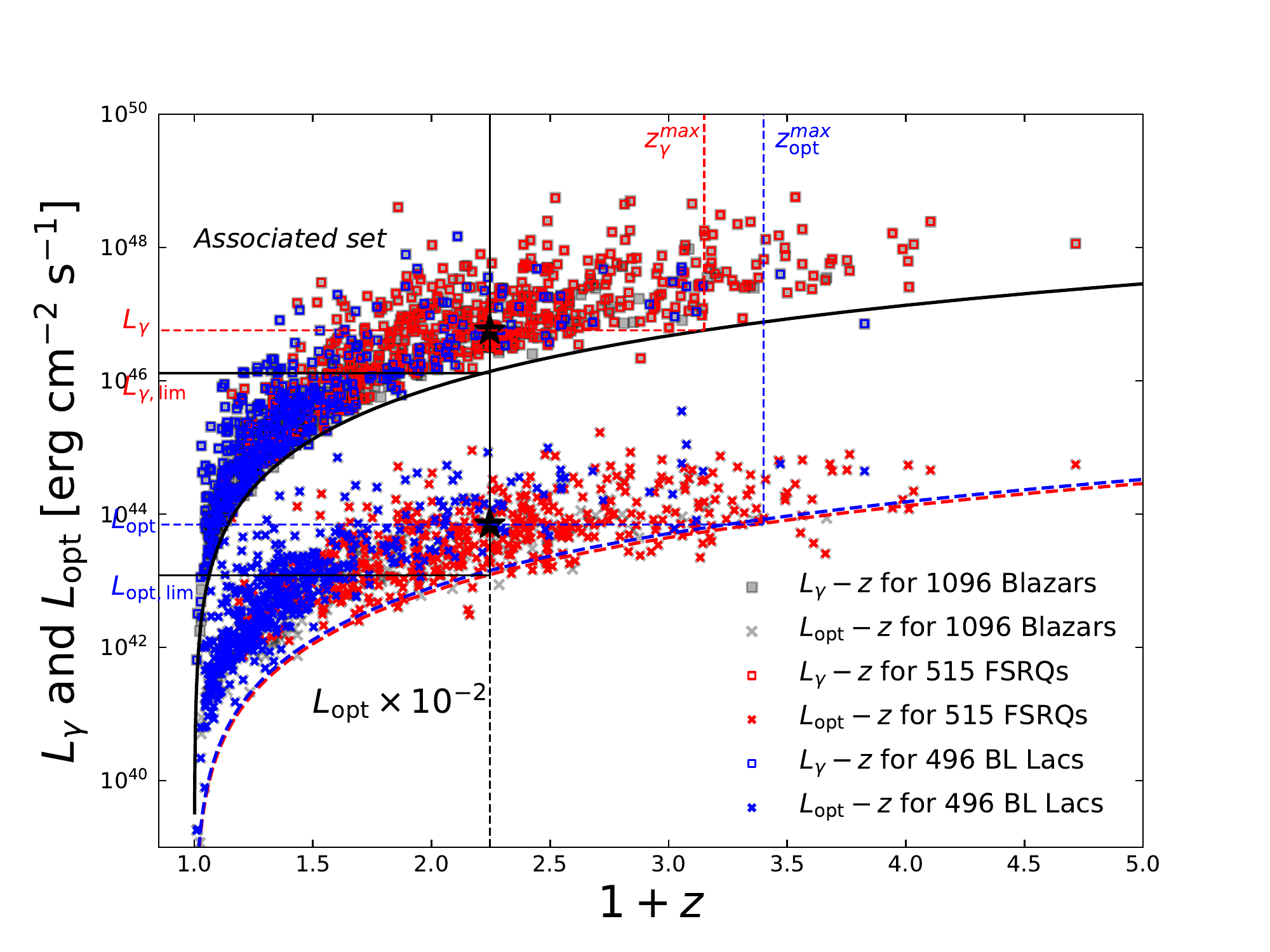}
\caption{The distributions of gamma-ray (squares) and optical (crosses) luminosities  vs redshift of 1124 blazars. The black, red and blue curves show the respective truncation boundaries assuming limiting (corrected) gamma-ray flux of $S^{\rm cr}_{\rm limit}=1.26
    \times 10^{-13}$ erg cm$^{-2}$ s$^{-1}$ with a fixed photon index $\Gamma = 2.24$ and optical flux of $F_{\rm opt, lim}=1.0 \times 10^{-13}$ erg cm$^{-2}$ s$^{-1}$  and $1.15 \times 10^{-13}$ erg cm$^{-2}$ s$^{-1}$ with a fixed spectral index $\alpha_{\rm opt} = 1.39$. For the source identified by the large black star, the redshift is shown by the vertical black dotted line, and gamma-ray and optical luninosities shown by the horizontal dashed lines, respectively. Intersection of these lines with the truncation curves gives the two maximum redshifts, $z_{\rm max}$, shown by vertical red and blue dotted lines. We classify this source as gamma-ray limited. The Associated sets for ranking in redshift are the sources inside the box defined by the black solid lines for gamma-ray and optical samples. We carry out similar calulation for FSQRs and BL Lacs, respectively.
}
\label{fig:3}
\end{figure}

Having calculated luminosities for blazars with known redshifts we show the luminosity-redshift scatter diagram of 1096 blazars in Figure \ref{fig:3} (separated into FSRQs (red) and BL Lacs (blue)) at gamma-ray range (upper, squares) and optical range (lower, crosses).
The limiting gamma-ray luminosity is shown by the solid black curve, and the limiting  optical luminosity $L_{\rm opt, min}(z)$ obtained from equation \ref{Lopt} with the above limiting flux and average index are shown by the red and blue curves  for FSRQs and BL Lacs, respectively.
Only sources above the truncation curves will be used in our analysis.

 To separate the sources into optically and gamma-ray limited subsamples, for each blazars with given $L_\gamma$ and $L_{\rm opt}$, we calculate the maximum redshifts that the source could have to be in the sample, (i.e. to be above the truncation curves) from the following two equations.
\beq
\label{zmax}
L_\gamma=4 \pi d_{L}^2(z_{\gamma, {\rm max}}) S^{\rm cr}_{\rm lim}K(z_{\gamma, {\rm max}})\,\,\,\,{\rm and}\,\,\,\,
L_{\rm opt} = 4 \pi d_{L}^2(z_{\rm opt, max}) F_{\rm opt, lim}  (1+z_{\rm opt, max})^{\alpha_{\rm opt}-1},
\eeq
as demonstrated for the source identified by a large black star in Figure \ref{fig:3}.
Using the limiting flux limits $S^{\rm cr}_{\rm{lim}}=1.26 \times 10^{-12}$ erg cm$^{-2}$ s$^{-1}$ and $F_{\rm opt, lim}=1.0 \times 10^{-13}$ erg cm$^{-2}$ s$^{-1}$, we find that 763 of blazars have $R=z_{\rm{opt,max}}/z_{\gamma,\rm{max}}>1$, thus  classified as gamma-ray limited, and the rest (325 blazars with $R<1$) are classified as optically limited. With the same limited flux, we obtain 274 of FSRQs with $R>1$ and 235 with $R<1$.
For BL Lacs, 428 sources have $R>1 $ with $F_{\rm opt, lim}=1.15 \times 10^{-13}$ erg cm$^{-2}$ s$^{-1}$ and the lowest of flux $S^{\rm cr}_{\rm{lim}}=1.40 \times 10^{-12}$ erg cm$^{-2}$ s$^{-1}$ and only 67 sources have $R<1$.
\footnote{Note that this separation can be obtained from the observed fluxes and magnitudes using the ratio $R^\prime={S^{\rm obs}_{\rm lim}\over S^{\rm obs}}\times 10^{0.4(G_{\rm lim}-G)}$, which ignoring small differences between the K-corrections is equal to $[d_L(z_{\rm opt, max}/d_L(z_{\gamma, {\rm max}}]^2$. Again $R^\prime > 1$ defines the gamma-ray limited sample.} Note that the ratio between the gamma-ray limited sample and optically limited are 2.35, 6.39 and 1.16 for all Blazars, BL Lacs and FSRQs, respectively. Indicating that more of BL Lacs are gamma-ray limited compared to FSRQs.
Because of this, the fact that the fraction of BL Lacs with no redshift is large, and the $p$-values given above for them is small, we do not carry out LF and co-moving density evolution determination for BL Lacs. This differences between BL Lacs and FSRQs could be due to high-luminosity and featureless optical non-thermal emission that swamps the light from the host galaxy making it difficult to obtain a redshift \citep[e.g.][]{2012MNRAS.422L..48P,2020arXiv200712661K}. In general, BL Lacs shows a complicated cosmological evolution; e.g~ HBL type BL Lacs have been shown to evolve less than the LBL types \citep[e.g.][]{2000AJ....120.1626R,2014ApJ...780...73A}.
Thus, our main focus will be the analysis of the evolution of gamma-ray LF of FSRQs.

\section{THE LUMINOSITY FUNCTION AND ITS EVOLUTION}
\label{LF}

In this section we calculate the luminosity evolution, LF and co-moving density evolution in
gamma-ray and optical ranges for FSRQs only, using the EP-L methods.

\subsection{Luminosity Evolution}
\label{LE}

\begin{figure}
\gridline{\fig{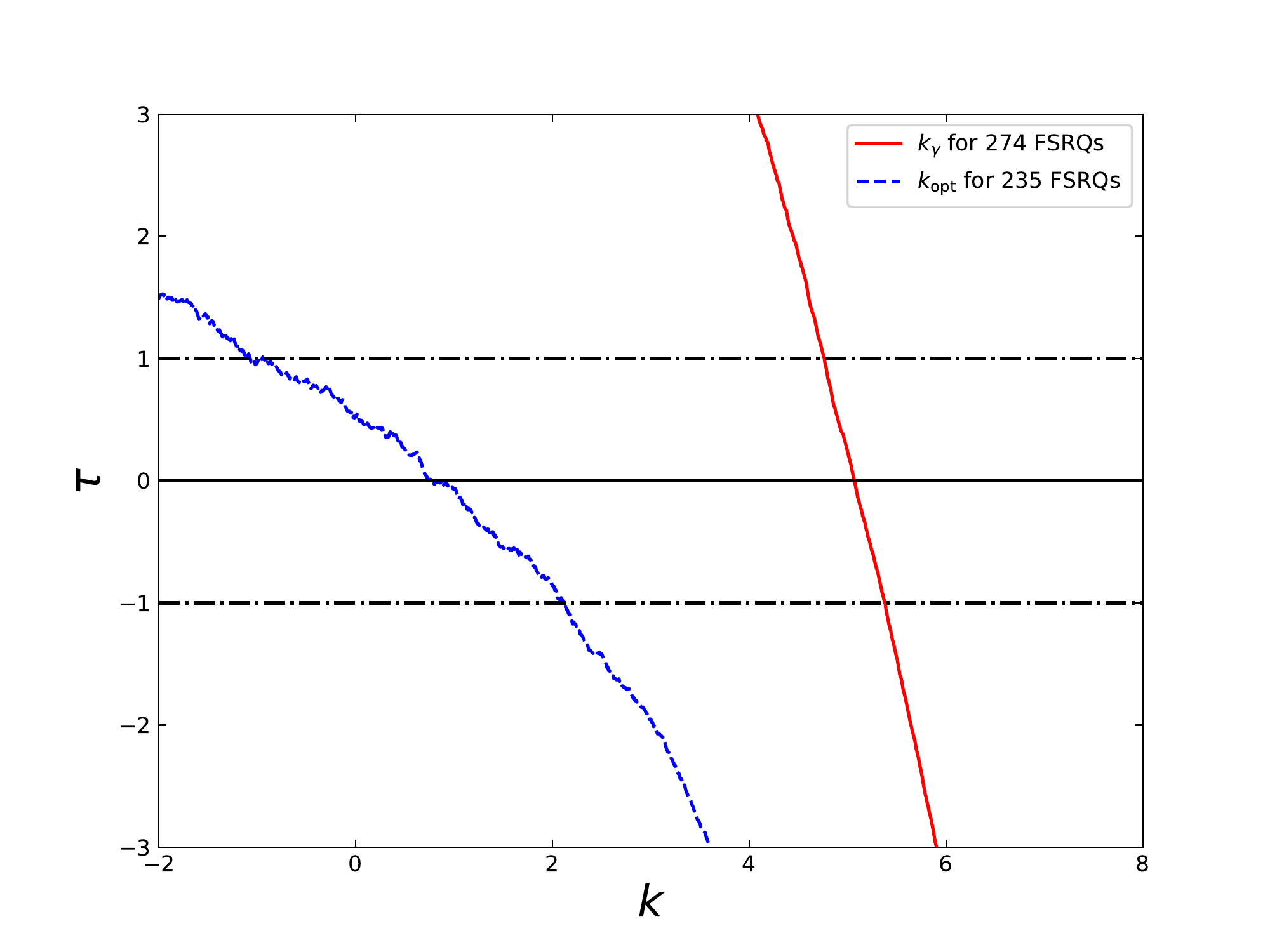}{0.4\textwidth}{(a)}
          \fig{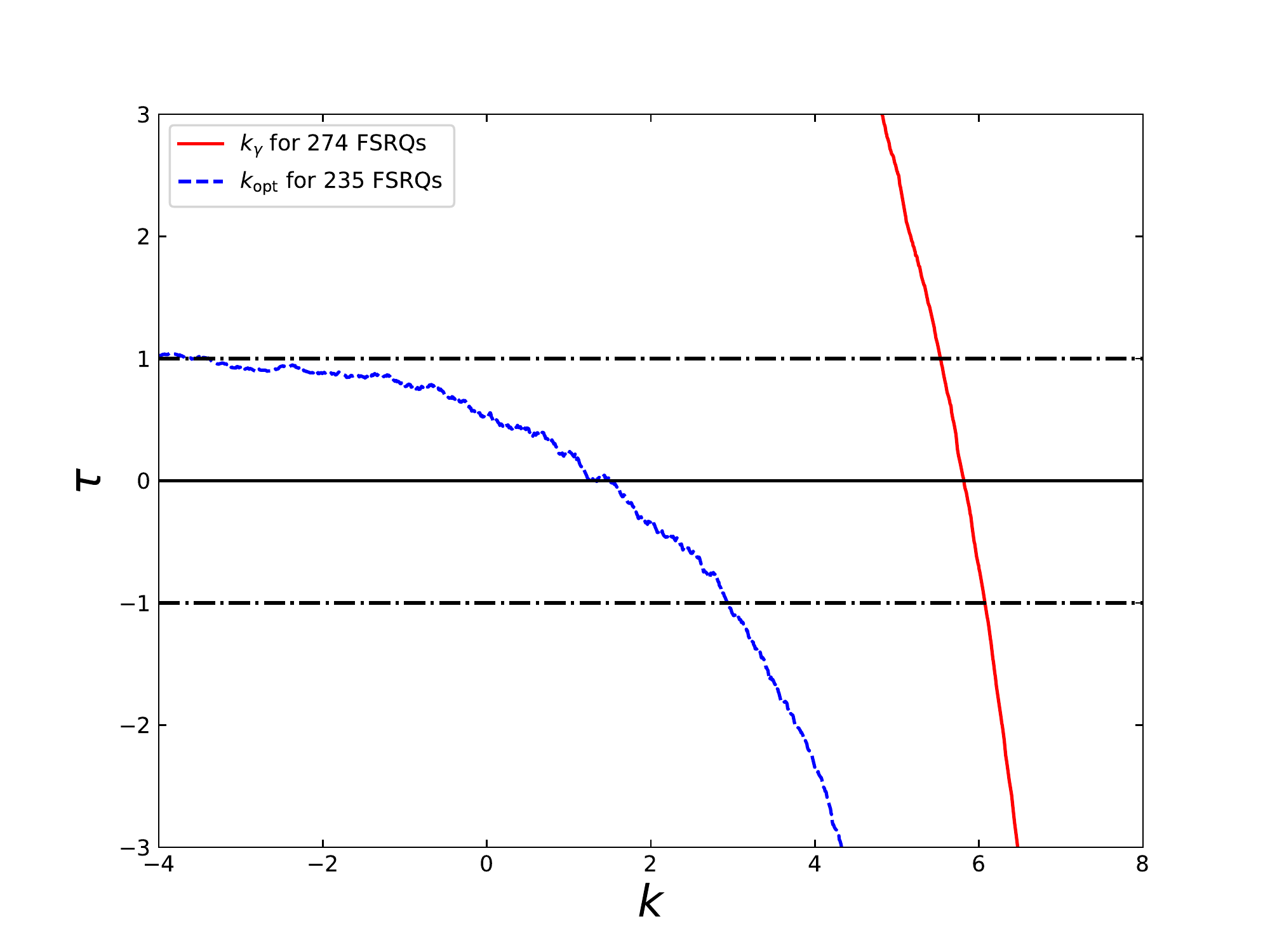}{0.4\textwidth}{(b)}}
\caption{Left (a): Test statistic $\tau$ vs evolution index $k$ for 274 gamma-ray limited (red solid line) and 235 optically limited (blue solid line) FSRQs for the luminosity
evolution function $g_i(z)=(1+z)^{k_i}$ yielding the best $\tau=0$ and one-sigma $\tau=\pm 1$ values of $k_{\gamma}=5.07^{+0.31}_{-0.31}$ and $k_{\rm opt}=0.77^{+1.34}_{-1.72}$. Right (b): Same as (a), but for luminosity evolution function
$g_{i}(z)=\frac{(1+z)^{k_i}}{1+(\frac{1+z}{Z_{cr}})^{k_{i}}}$ with $Z_{cr}=3.5$, and
results $k_{\gamma}=5.81 ^{+0.26}_{-0.28}$ and $k_{\rm opt}=1.50^{+1.43}_{-4.90}$.
}
\label{fig:kappatau}
\end{figure}

As described above the first step to this end is the test of independence of luminosity and redshift, and the determination of a modified luminosity (corrected for the luminosity evolution described by a functions $g_i(z)$) that is independence of (or is uncorrelated with) redshift. As evident from Figure \ref{fig:3} the luminosities show strong correlation with redshift most of which is due to the bias introduced by the data truncations. The EP method accounts for this using rank statistics based on a modified version of the Kendall tau test;
$\tau=\Sigma_i(R_i-E_i)/\sqrt{\Sigma_iV_i}$, where $R_i$ are the normalized ranks of the sources in their {\it associated sets} (consisting of sources with $z_j<z_i$ and $L_j>L_{i,{\rm lim}}$, for ranking in $L$).
More details description of this test can be found in
\cite{1992ApJ...399..345E,1999ApJ...518...32M,2014ApJ...786..109S,2015ApJ...806...44P}, For the null hypothesis of independence (i.e.~no luminosity evolution) the expected ranks and variances are $E_i=1/2$  and $V_i=1/12$. We find that for both subsamples  $\tau(k=0)>5$ rejecting the null hypothesis at $>5\sigma$ level. To  obtain a measure of the correlation or the luminosity evolution function we use both a simple power-law luminosity evolution functions, $g_i(z)=(1+z)^{k_i}$, or, as shown in \cite{2013ApJ...764...43S}, a more realistic (but still single parameter) form with a flattening above the critical redshift $Z_{cr}\sim 3.5$ that accounts for decreasing cosmic expansion time with increasing redshift:
\begin{equation}
\label{gofz}
g_i(z)=\frac{(1+z)^{k_i}}{1+(\frac{1+z}{Z_{cr}})^{k_i}}.
\end{equation}
Figure \ref{fig:kappatau} shows the variation of $\tau$ with evolution indicies $k_\gamma$ and $k_{\rm opt}$ showing the best-fit values and 1-sigma ranges of  $k_{\gamma}=5.07^{+0.31}_{-0.31}$ and $k_{\rm opt}=0.77^{+1.34}_{-1.72}$, for the simple power law evolution, and $k_{\gamma}=5.81 ^{+0.26}_{-0.28}$ and $k_{\rm opt}=1.50^{+1.43}_{-4.90}$ for the form given by Equation (\ref{gofz}). Here we are interested mainly on gamma-ray luminosity evolution but we have also evaluated the optical luminosity evolution for the 
optically limited sample to test the accuracy of the separation technique used here. Considering the sample size the results agree with the optical luminosity  evolution derived for a  much larger SDSS sample of quasars, $k_{\rm opt}\sim 3.5\pm 0.3$ \citep{2013ApJ...764...43S} and $k_{\rm opt}\sim 4.4$ used in \cite{2015IAUS..313..333P}. Furthermore, the gamma-ray evolution index $k_\gamma\sim 5.81$ is closer to the evolution index at radio wavelengths, $k_{\rm rad}\sim 5.5\pm 0.2$.

\subsection{Gamma-ray and Optical Luminosity Correlation}
\label{sec:llcorr}

\begin{figure}[ht!]
\plotone{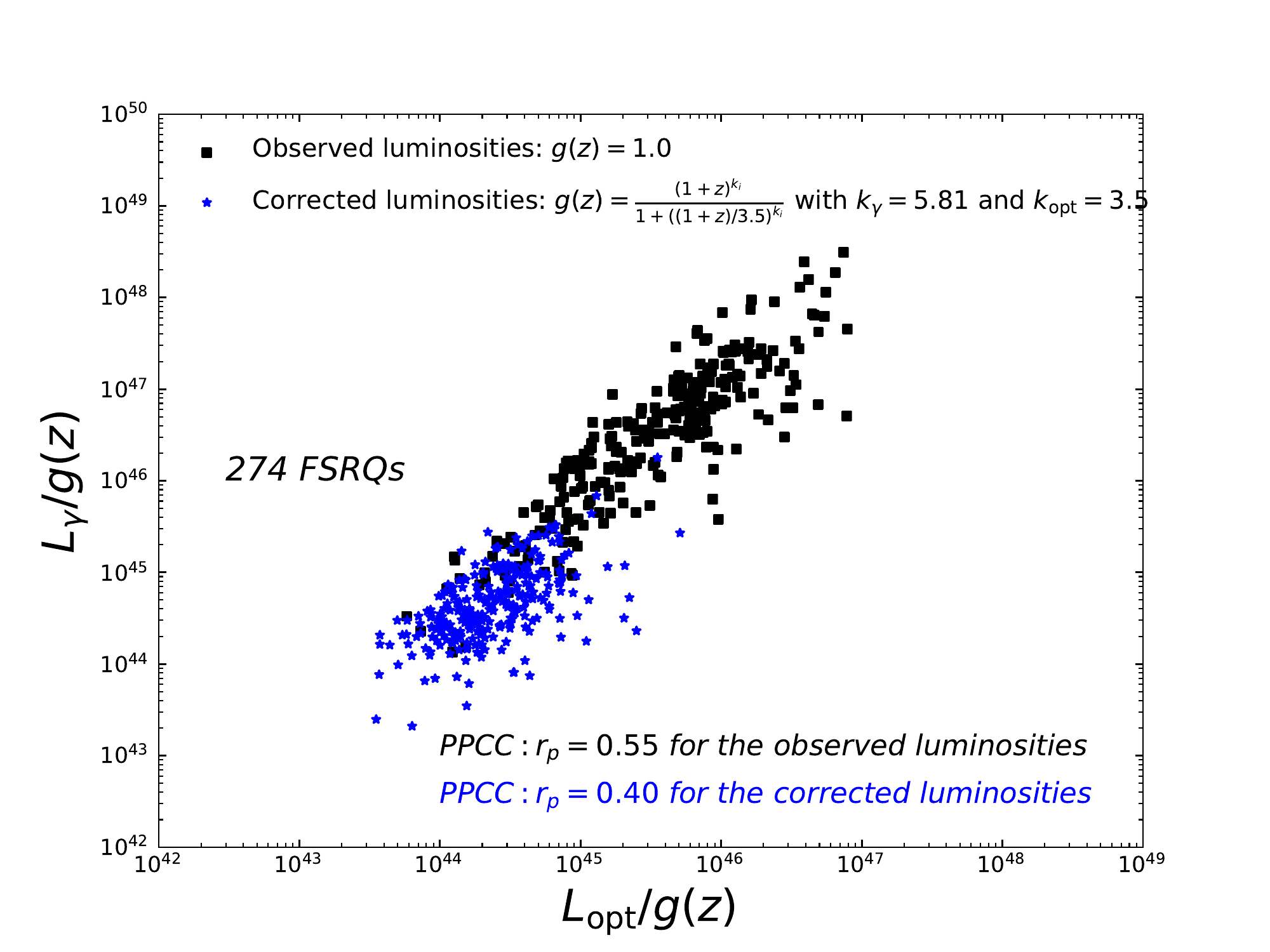}
\caption{The black points show the {\it observed}  gamma-ray luminosity vs  optical luminosity  of {\it gamma-ray limited sample} of 274 FSRQs, showing a strong correlation due to their similar dependence on the luminosity distance (or redshift) and similar luminosity evolutions. The blue points show the local gamma-ray luminosity vs the local optical luminosity obtained using the evolution function given in Eq. \ref{gofz} with $k_\gamma=5.81$ and $k_{\rm opt}=3.5$. The local luminosities also show significant residual correlation, not all of which is intrinsic; some is due to similar dependence on luminosity distance. }
\label{fig:lumcorr}
\end{figure}

As mentioned by \cite{2013ApJ...764...43S}, and can be seen in Figure \ref{fig:lumcorr}, the observed luminosities show a strong correlation mostly because of (i) their almost identical dependence on the luminosity distance, except for a minor difference in
the K-correction terms,  and (ii) the strong and  similar luminosity evolutions described above. As shown in \cite{2015IAUS..313..333P}, we can correct for the latter effect by using the de-evolved, or (when $g(z=0)=1$ which is the case here) the local luminosities $L_{i,0}=L_i/g_i(z)$, and use {\it partial (Pearson or Kendall)} correlation coefficients to account for the former.

Figure \ref{fig:lumcorr} shows scatter diagram of the observer (black points) and local or de-evolved (blue points) gamma-ray vs optical luminosity scatter diagrams  of the gamma-ray limited FSRQs samples, using the results from evolution function given in Equation (\ref{gofz}), and the more accurate optical evolution index $k_{\rm opt}=3.5$ obtained from the larger SDSS optically selected sample.
As evident the bulk of the observed correlation is due to similar luminosity evolutions, but a considerable residual correlation is still present in the evolution corrected luminosities. 

Following \cite{2015IAUS..313..333P}
we quantify the correlation between the local luminosities by defining a correlation corrected gamma-ray (local) luminosity
\beq
\label{eq:llcorr}
L^{cr}_{\gamma,0}=L_{\gamma,0}(L_0/L_{{\rm opt},0})^{\alpha_\gamma},
\eeq
where $L_0$ is a fiducial luminosity (here $L_0=10^{45}$ erg/s) and $L_{{\rm opt},0}=L_{\rm opt}/g_{\rm opt}(z)$ with $k_{\rm opt}=3.5$, and calculate the partial Pearson correlation coefficient (PPCC) as a function of $\alpha_\gamma$.

\begin{figure}[ht!]
\plotone{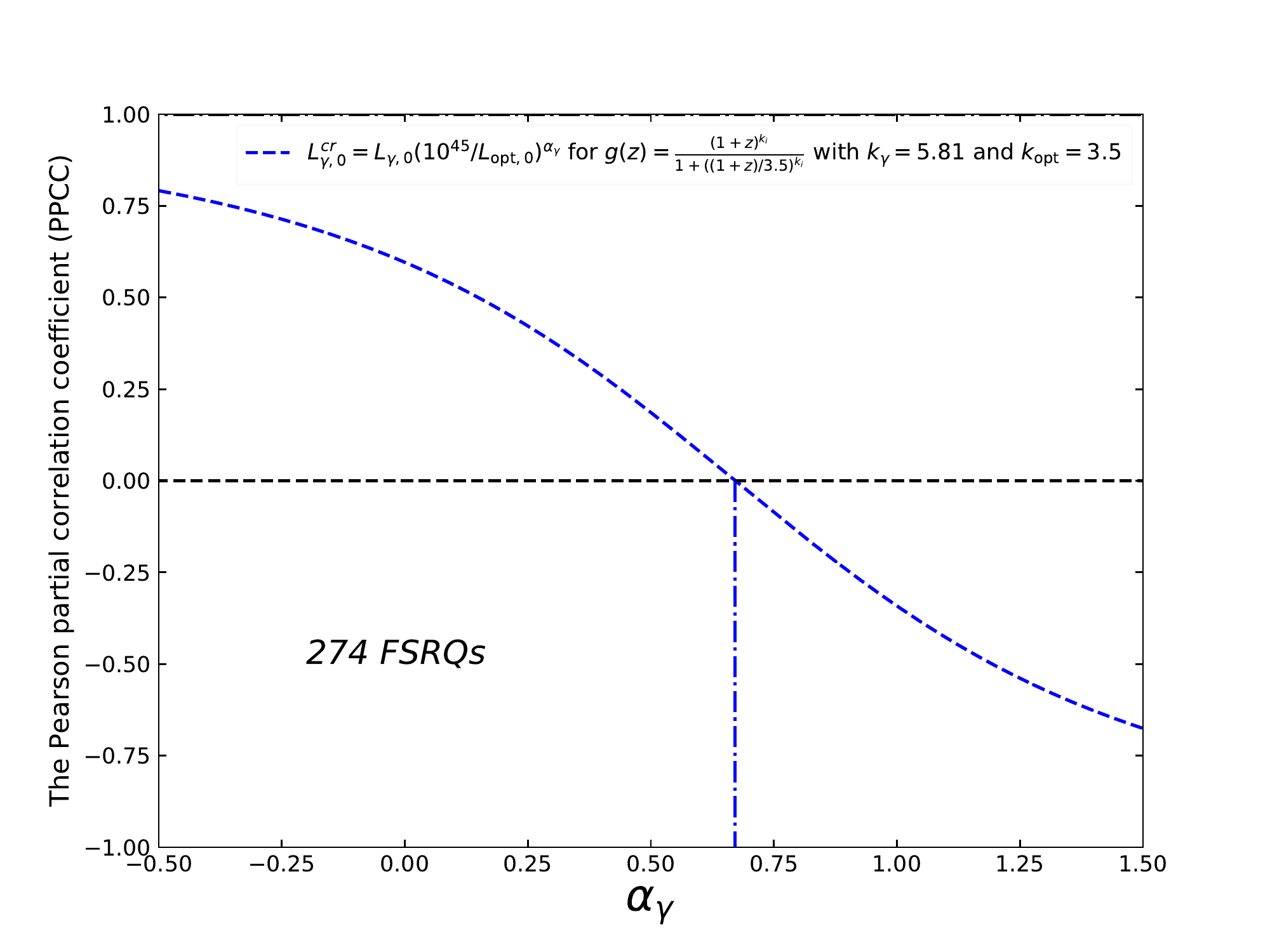}
\caption{Variation of the partial Pearson correlation coefficient (PPCC) with the $L-L$ correlation index $\alpha_\gamma$ (defined in Eq. \ref{eq:llcorr}) giving  best values and one sigma ranges of $\alpha_\gamma=0.679 \pm 0.043$ for FSRQs. }
\label{fig:lumbeta}
\end{figure}

Figure \ref{fig:lumbeta} shows the variation of the $PPCC$ statistic with the index $\alpha_\gamma$ using the evolution function of Eq. (\ref{gofz}), yielding the value of $\alpha_\gamma$ that give zero partial correlation coefficient and its 1-$\sigma$ range: $0.68 \pm 0.04$ for FSRQs. 
This is to be compared with  the index $\alpha_{\rm rad}=0.325\pm 0.050$
found by \cite{2015IAUS..313..333P}
for radio-optical $L-L$ correlation and  $\alpha_{\rm mIR}=0.75\pm 0.1$ for mid-infrared-optical $L-L$ correlation found by \citet{2019ApJ...877...63S}.

\subsection{Gamma-ray Luminosity Function and Density Evolution}
\label{LFgamma}

Having established the independence of the local luminosity and redshift we use the EP-L method to obtain their mono-variate luminosity  and redshift distributions [the LF, $\psi(L_0)$, and the co-moving density evolution, $\rho(z)$]. This method (see, e.g.~\citet{2015ApJ...806...44P}) gives a nonparametric
and point-by-point description of the cumulative distributions \tcr{}{}(traditionally in descending order in luminosity and ascending order in redshift)
\begin{equation}
\label{cumdist}
\phi(L_0)= \int_{L_0}^{\infty} \psi(L^{\prime}) dL^{\prime}\,\,\,\,{\rm and} \,\,\, \sigma(z)=\Omega\int_0^z \frac{dV}{dz^\prime}\rho(z^\prime)dz^\prime,
\end{equation}
as  histograms:
\begin{equation}
\label{associated}
\phi(L_{0,i})=\prod_{j=1}^{i}\left(1+\frac{1}{N_j}\right) \,\,\,\,{\rm and}\,\,\,\, \sigma(z)=\prod_{j=1}^{i}\left(1+\frac{1}{M_j}\right),
\end{equation}
where $N_j$ is the number of object in the associated set of
object $i$ (with variables $L_i, z_i$) consisting of set of sources with $L_j>L_i$, and $z_{j}<z_{{\rm max},i}$ (or $L_{{\rm min},j}<L_i$),
and $M_j$ is the number of object in the associated set of
object $i$  consisting  of sources with $z_j<z_i$, and $L_j>L_{{\rm min},i}$ (or $z_{{\rm max},j}>z_i$.). In Eq.~(\ref{cumdist}) {\bf $\Omega=10.38$ sr} is the the total
sky coverage, with $|b| > 10^o$.

\begin{figure}
\gridline{\fig{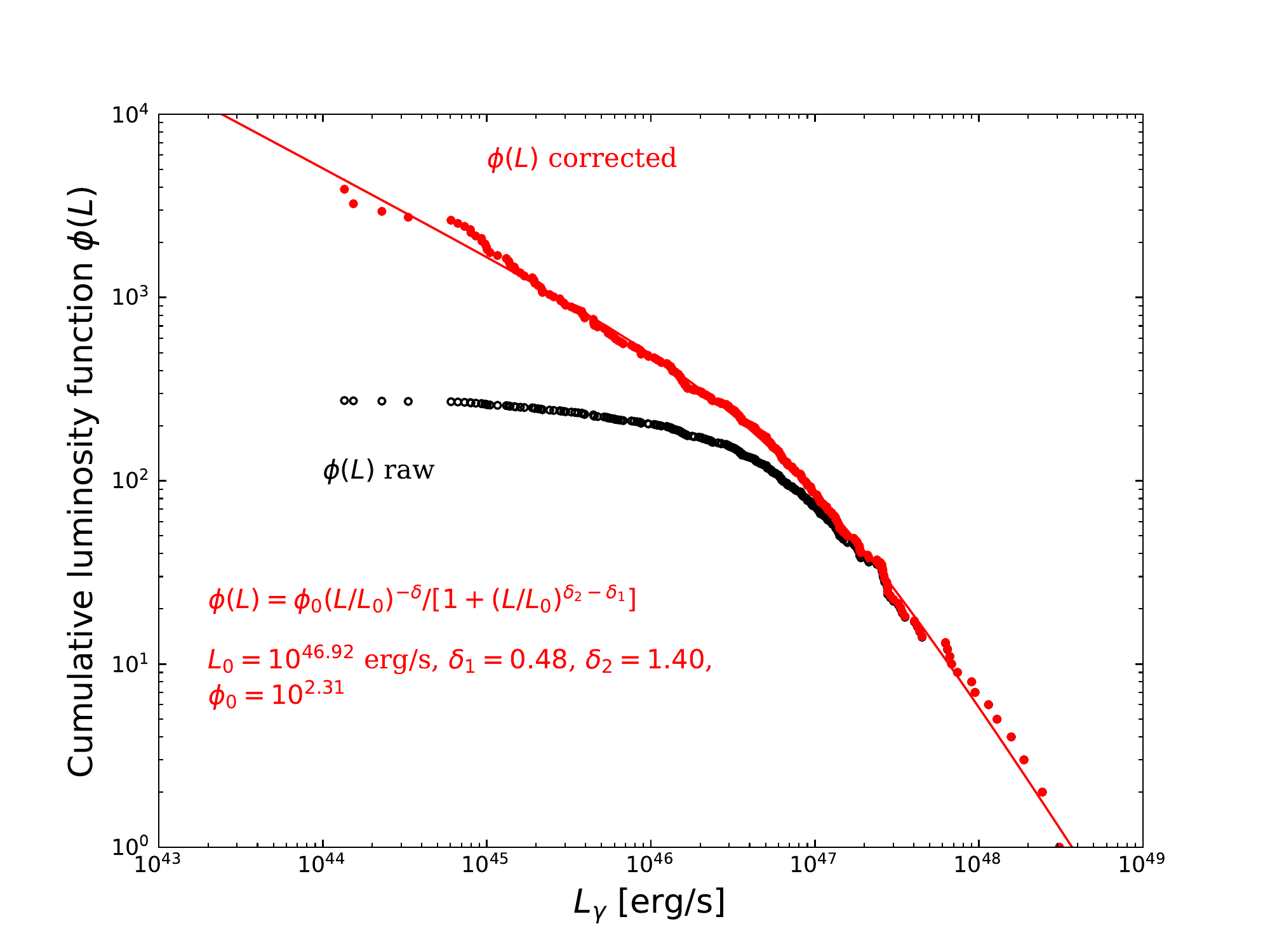}{0.4\textwidth}{}
          \fig{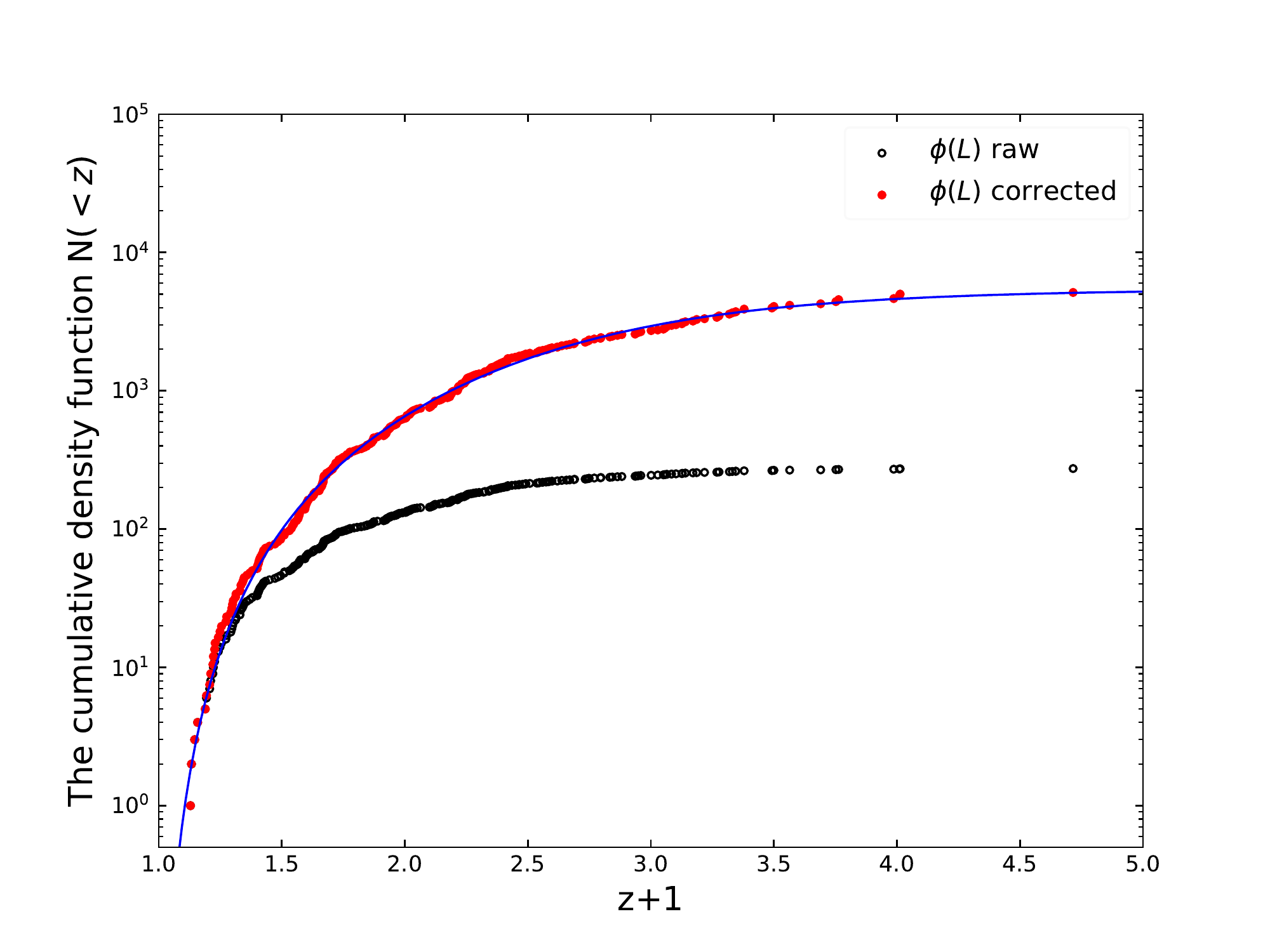}{0.4\textwidth}{}
          }
\caption{(Left): The cumulative local gamma-ray LF of FSRQs, $\Phi(L_0)$, corrected for the gamma-ray flux limit and luminosity evolution (red points) compared with raw histogram of local luminosities, $N(>L_0)$, (black points). The corrected LF is fit to a broken power law.  (red curve with specified parameters in red).
(Right): The cumulative density distribution $\sigma(z)$ vs. redshift corrected for luminosity evolution and selection effects (red points) compared with raw cumulative observed redshifts, $N(>z)$, (black points). The density evolution $\sigma(z)$ can be fit quite well with redshift integral (Eq.~(\ref{cumdist})) for the co-moving density $\rho(z)$ given in Eq.~(\ref{sigmafit}); (blue curve).
}
\label{fig:cumdist}
\end{figure}

Figure \ref{fig:cumdist} shows the histograms for the cumulative luminosity and density distributions corrected for the gamma-ray flux limit and the corresponding raw cumulative distributions of $L_0$ and $z$, $N(>L_0)$ and  $N(<z)$.

\begin{figure}
\gridline{\fig{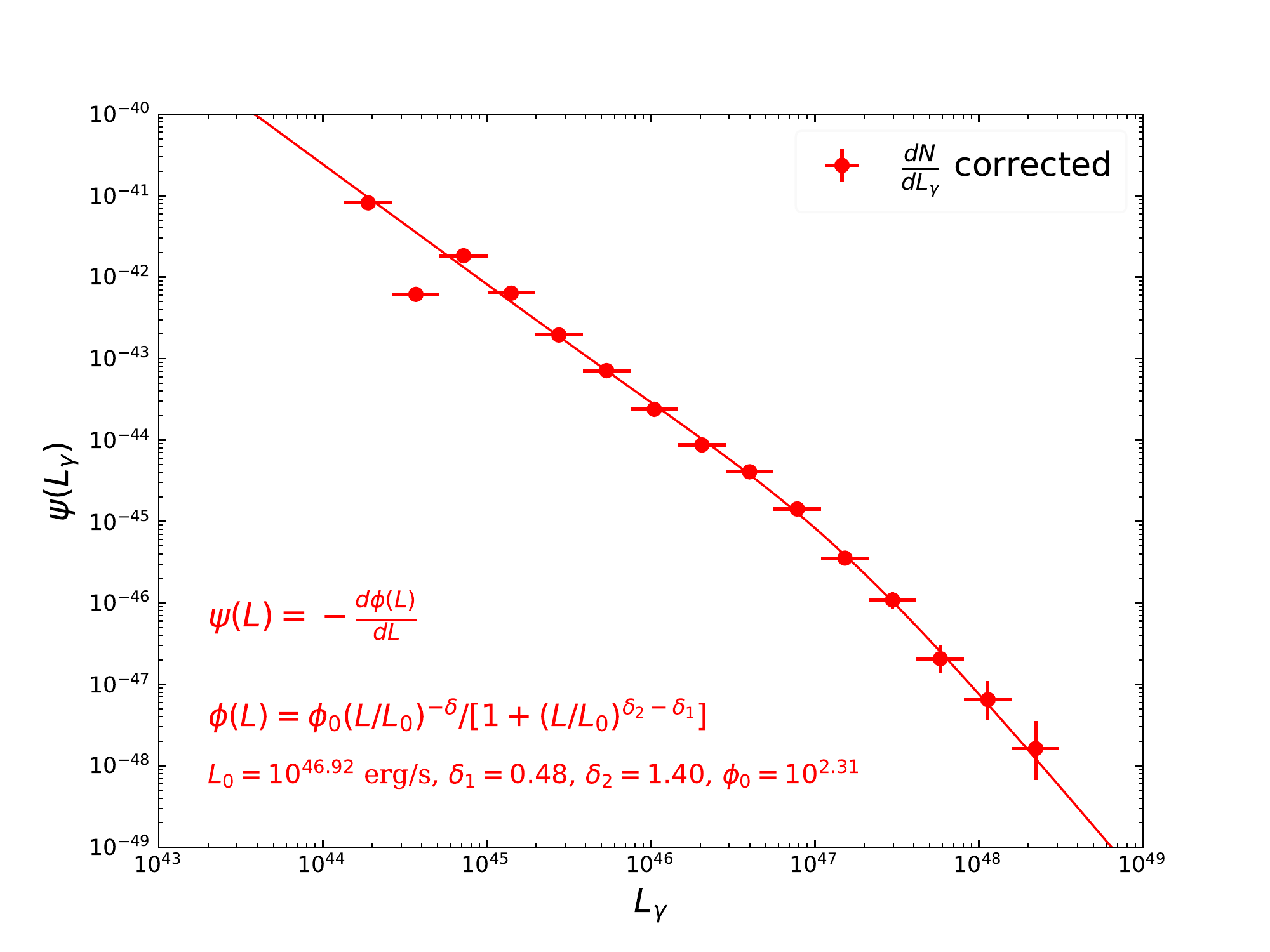}{0.4\textwidth}{}
          \fig{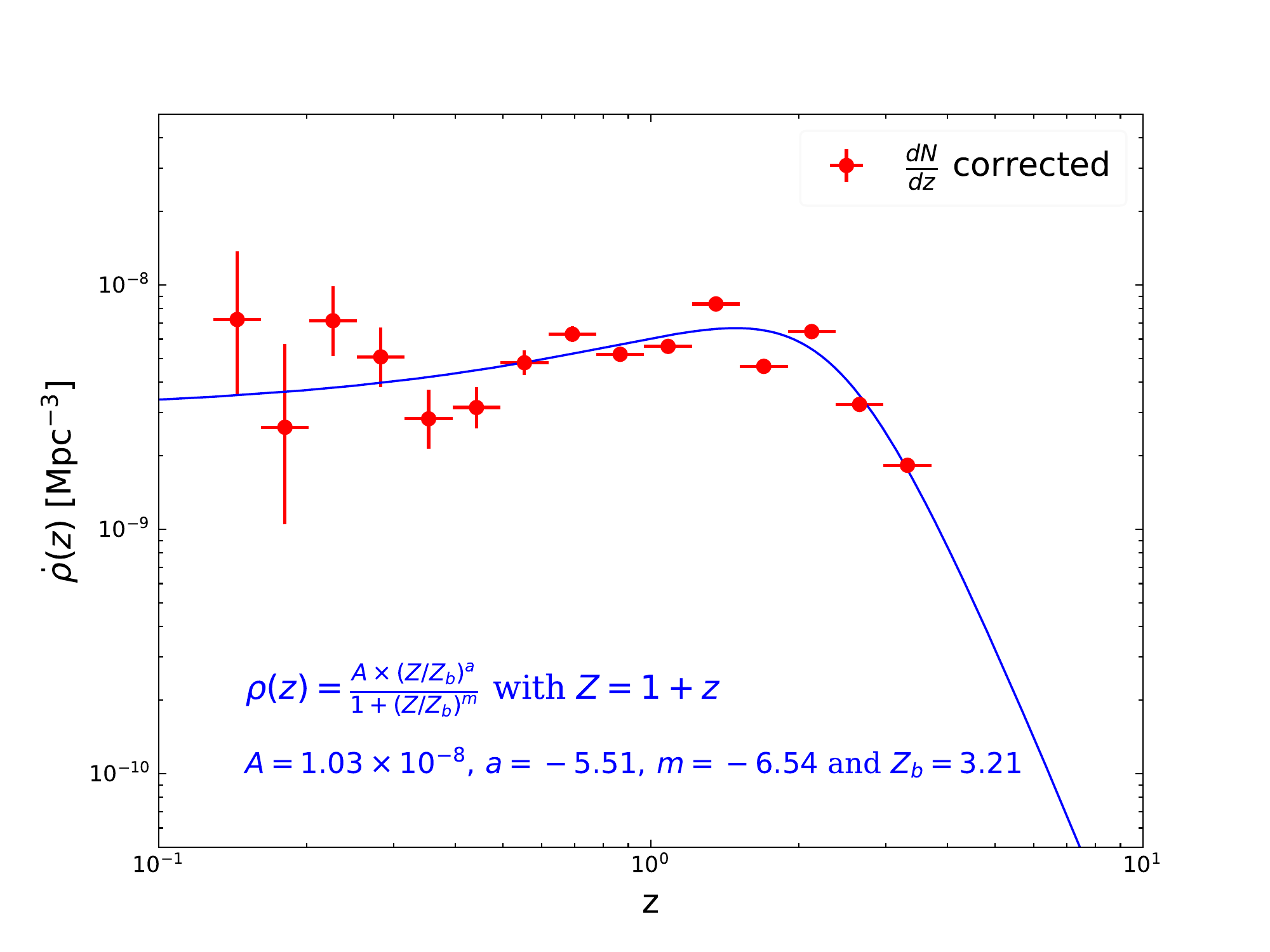}{0.4\textwidth}{}
          }
\caption{The differential local gamma-ray LF (left) and co-moving density evolution (right)
for gamma-ray limited FSRQs.
The red filled points are obtained non-parametrically and directly from differentiation of Eq. (\ref{associated}) smoothed by 15 point binned forms of Equations (\ref{diffdist}) showing some small scale variations. The curves are obtained from the differentiation of the  the broken power-law fit to the cumulative LF, yielding indexes 1.48 and 2.40 below and above the break $L_{\rm 0}= 8.3 \times 10^{46}$ erg s$^{-1}$, and the parametric analytic form for the co-moving density $\rho(z)$ given in Eq.~(\ref{sigmafit}) for parameter values $A=1.03 \times 10^{-8},\, a=-5.51,\, m=-6.54$ and $Z_b=3.21$.}
\label{fig:diffdist}
\end{figure}

The derivatives of the cumulative distributions give the differential distributions
\beq
\label{diffdist}
\psi(L_0)=-{d\phi(L_0)\over dL_0}, \,\,\,\, {\rm and} \,\,\,\, \rho(z) = \frac{d\sigma(z)}{dz} \times \frac{1}{\Omega dV/dz}.
\eeq
These  distributions can also be obtained directly and non-parametrically by the differentiation of the numerical histograms of Eq. (\ref{associated}) as described in  Appendix B of \cite{2015ApJ...806...44P}. A point by point derivative, however, yields a noisy result, which can be improved by smoothing. Using a 15 point bin interpolation smoothing, we obtain the differential distributions shown by the points with error bars in Figure \ref{fig:diffdist}.
Alternatively, one can fit the cumulative distributions in Figure~\ref{fig:cumdist} to analytic forms, which can then easily be differentiated to obtain the differential distributions.
The cumulative corrected local LF can be fit to a smooth broken power law (with parameters shown on the left panel of Figure~\ref{fig:cumdist}). Differentiating this we obtain the differential LF shown by the red curve in Figure \ref{fig:diffdist}, which  also has a smooth broken power law form with power law indexes 1.48 and 2.40  below and above the break luminosity $L_{0}=8.3 \times 10^{46}$ erg s$^{-1}$, which  agrees well with the 15 bin interpolation smoothing non-parametric results (red points).

For obtaining the  an analytic form for the density evolution we use a slightly different procedure. We start with the  parametric broken power law form

\begin{equation}
\label{sigmafit}
\rho(z)= \frac{A\times (Z/Z_b)^a}{1+(Z/Z_b)^{m}} \,\,\,\, {\rm with} \,\,\,\,  Z=1+z,
\end{equation}
integrate it using the right portion of Eq. (\ref{cumdist}), and fit to the histogram of $\sigma(z)$ to find the best values for the four parameters for the above analytic form ($A=1.03 \times 10^{-8}, \, a=-5.51,\, m=-6.54$ and $Z_b=3.21$) shown by the blue curve
on the right panel of  Figure~\ref{fig:diffdist}. which agrees well with the 15 bin interpolation smoothing non-parametric results (red points).

Again as expected, the indexes of the cumulative and differential LF differ by one. For FSRQs, our result show a larger (steeper) high luminosity index compared to the results of \cite{2012ApJ...751..108A}, obtained by forward fitting method assuming a pure luminosity evolution.
The co-moving density evolution of FSRQs peaks at around redshift $\sim 2.0$, which is slightly large than that of SDSS optical-radio quasars found by
\cite{2014ApJ...786..109S} roughly agrees with the fact that the cosmological sources have
their redshift evolution peaks at $z \sim 1-2$ as does AGN activity in
general \cite[e.g.][]{2003ApJ...598..886U}.

\section{Contribution of Blazars to EGB}
\label{egb}

Figure \ref{fig:ScrGamdist} shows the Extragalactic portion of the diffuse gamma-ray energy intensity (EGB)  above $E_0=100$ MeV observed by Fermi-LAT \citep{2015ApJ...799...86A}. What is shown is $E^2dN/dE$, where $dN/dE$ is the specific photon number intensity (in ph cm$^{-2}$ s$^{-1}$ sr$^{-1}$ MeV$^{-1}$). To avoid confusion with other usage of $N$ we use more common symbols for intensity ${\cal I}(E)=dN/dE$ and use $I(E)=E\times dN/dE$, for specific energy intensity and ${\cal E}(E)=E^2dN/dE$ as a measure of total energy flux per steradian. The question we like to address here is what fraction of this diffuse EGB is due to population of all blazars. A common practice to this end is to use multi-dimensional integration (over redshift, luminosity, spectral characteristics) of the luminosity function and its various evolutions (luminosity and co-moving density) to estimate the amplitude and spectra of these intensities. As we have shown in \S 3, all of these information is obtained from {\it the flux and spectral measurements} of a well defined observed sample of sources with redshift (described in \S 2). Thus, such a procedure seems a somewhat circular approach which can amplify the observational errors, and is subject to completion uncertainty of samples with redshift. Such completeness issues do not affect the observed flux distributions and its limit. Thus, we use a procedure similar to the one used in \cite{2012ApJ...753...45S}, which is a simple method of obtaining the amplitudes of the above intensities directly from the flux and spectral index distributions described in \S 2, without requiring any knowledge of redshifts or recourse to the LF.

\begin{figure}
\plottwo{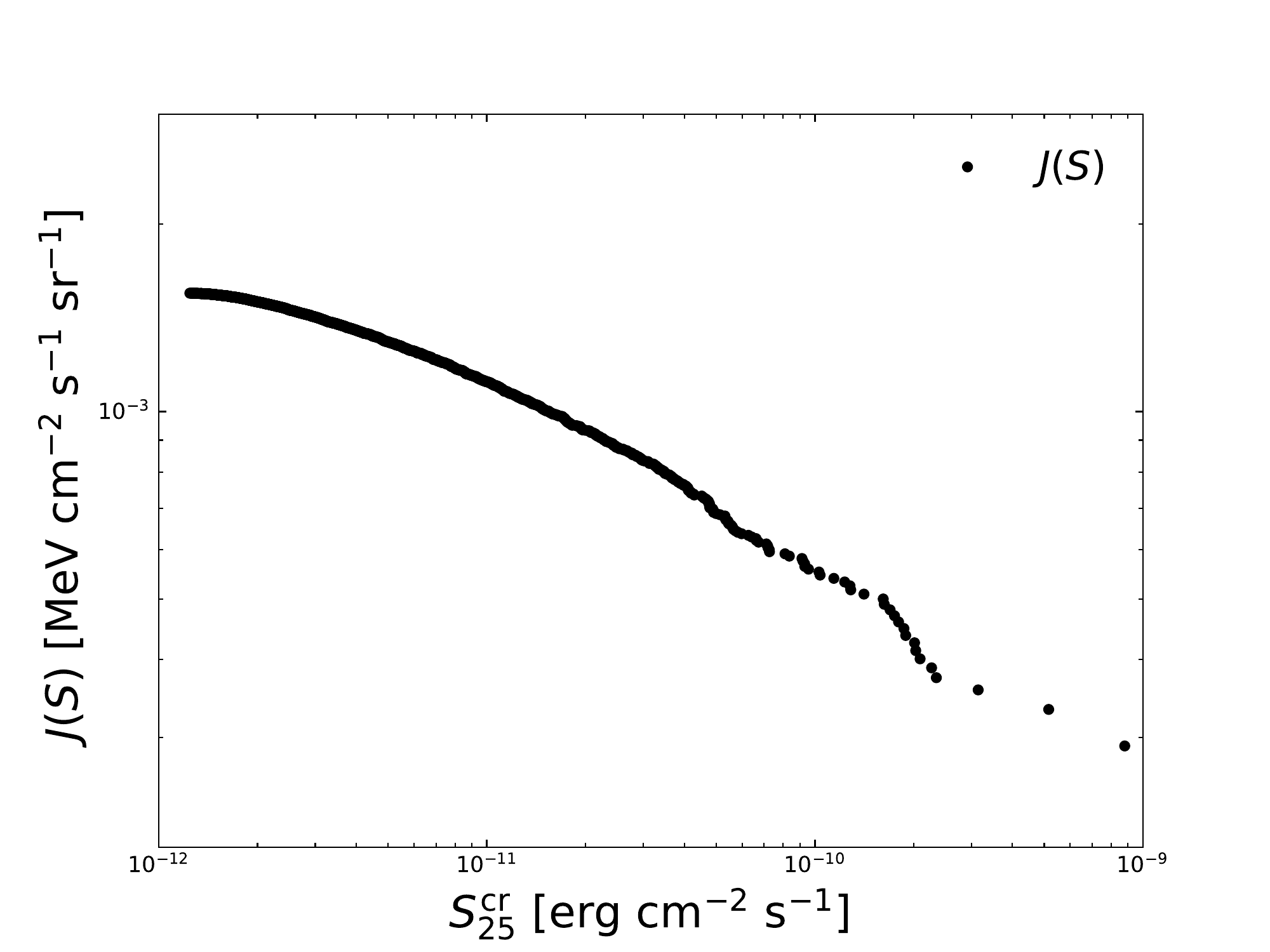}{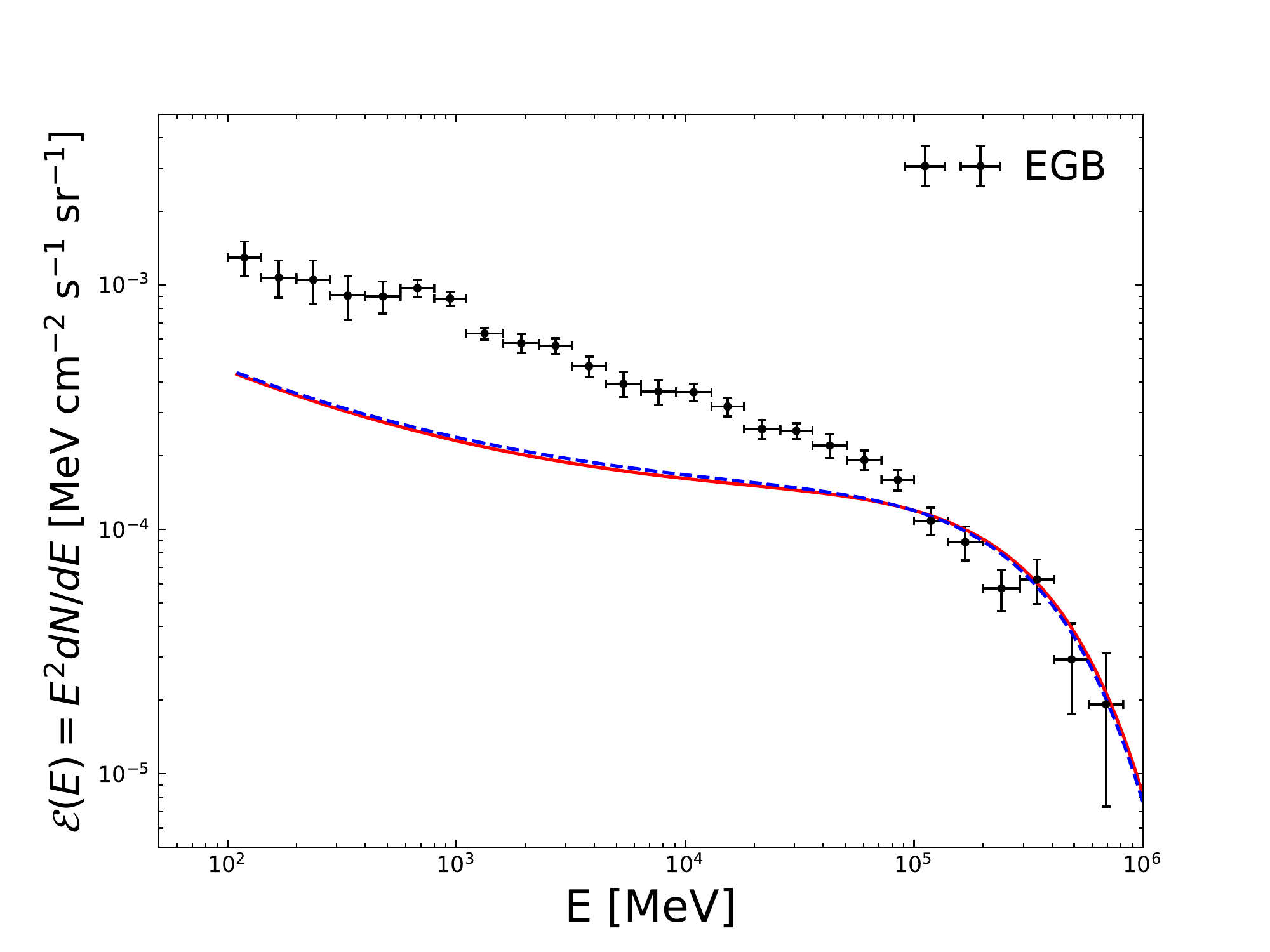}
\caption{{\bf Left:} J(S) is obtained by only the non-parametrically determined  cumulative  distribution of $N(>S)$ shown on the left panel of Figure  2.
 {\bf Right:} ${\cal I}(>S)$ calculated with $\Gamma_0=2.18$ and $\sigma=0.28$ are predicted contribution of Fermi-LAT blazars to EGB spectrum from Ackerman et al. (2015). Note red solid line is obtained from numerical integration of Eq.~(\ref{HofE2}), and the dashed blue line is based on the analytic expression  Eq.~(\ref{HofE5}).
 }
\label{fig:EGB}
\end{figure}

In section 2, After relatively small correction for truncation we obtained a bi-variate distribution $G(S, \Gamma)$ (where for convenience we have dropped the subscript 25 and super script $cr$ from the energy flux $S$), which because of the independence of $S$ and $\Gamma$ can be written as
\begin{equation}
\label{Gofsg}
G(S,\Gamma) = dN/dS \, \times \,  h(\Gamma).
\end{equation}
The two terms here are shown in the middle and right panel of Figure \ref{fig:ScrGamdist}, and the cumulative distribution, $N(S)$, the so called log $N$-Log $S$ relation (when $\Gamma$ distribution is normalized as $\int_{-\infty}^\infty h(\Gamma) d\Gamma=1$) is shown on the left.
The contribution to the energy and  photon intensity of EGB can be calculated as
\begin{equation}
\label{egbE}
I_\gamma(>S)=\int_S^\infty S'(dN/dS')dS'\, \int_{-\infty}^\infty h(\Gamma) d\Gamma,
\end{equation}
and
\begin{equation}
\label{egbN}
{\cal I}_\gamma(>S)=\int_S^\infty (dN/dS')dS'\, \int_{-\infty}^\infty F(S',\Gamma)\times h(\Gamma) d\Gamma \end{equation}
where according to  Eq. \ref{eq:S-F}, the photon number flux $F=(S/E_1)g(\Gamma)$.
In both cases the integrating over $S$ by part (as in \cite{2014ApJ...786..109S}) we obtain
\begin{equation}
\label{JofS}
 J(S)=\int_S^\infty S'(dN/dS') dS'=SN(>S)+\int_S^\infty N(>S')dS',
\end{equation}
shown on the left panel of Figure \ref{fig:EGB}, involving only the non-parametrically determined cumulative distribution  $N(S)$ shown on the left panel of Figure \ref{fig:ScrGamdist}.

We thus can calculate these intensities up to the limiting flux $S_{\rm lim}=1.2\times 10^{-12}$. This gives 
$J(S_{\rm lim})=1.6 \times 10^{-3}$ MeV cm$^{-2}$ s$^{-1}$ sr$^{-1}$,
and upon numerical integration over $\Gamma$ using the observed distribution (Fig.~2, left) we obtain ${\cal I}_\gamma(>S_{\rm lim})=3.4 \times 10^{-6}$  cm$^{-2}$ s$^{-1}$ sr$^{-1}$, which is about $31$\% of the observed EGB photon intensity \footnote{ The total sky intensity of $1.1 \times 10^{-5}$ ph cm$^{-2}$ s$^{-1}$ sr$^{-1}$, which includes the isotropic diffuse gamma-ray background
(IGRB) plus detected sources \citep{2015ApJ...799...86A}}.

We can similarly calculate the {\it spectrum} of EGB and compare it with observations shown in Figure \ref{fig:EGB}(right). The total energy spectrum is obtained from
\begin{equation}
\label{egb1}
{\cal E}_\gamma(E)(>S)=E^2dN/dE=E^2 \int_S^\infty dS'\, \int_{-\infty}^\infty d\Gamma  f_\Gamma(E)
G(S', \Gamma),
\end{equation}
for which we need  the form, $f_\Gamma(E)$, of the spectral distribution of blazars. In the 100 MeV to 100 GeV range most blazars show a power law spectrum but as evident from Figure \ref{fig:EGB}, the spectrum must turn over at higher energies. Thus, we assume a spectrum consisting of a power law with
exponential cutoff,  $f_\Gamma(E)=f_0\times (E/E_1)^{-\Gamma}e^{-E/E_c}$, with $E_c=300$ GeV. Ignoring the $\sim 30$\% effect of the cutoff in the $E_1=$100 MeV to 100 GeV range, we can write $S=E_1^2(f_0/g(\Gamma))$, where
\begin{equation}
\label{fo}
g(\Gamma)=(\Gamma-2)/(1-10^{3(2-\Gamma)})
\end{equation}
except $g(2)=1/\ln 10^3\sim 0.14$.
Substitution of this in Eq. (\ref{egb1}), and with  the help of the derivation given above we obtain

\begin{equation}
\label{egb2}
{\cal E}_\gamma(E)(>S)=E^2 H(E)J(S)\,\,\,\, {\rm  with}\,\,\,\,
H(E)=\frac{e^{-E/E_c}}{E_1^2}\int_{-\infty}^\infty h(\Gamma)g(\Gamma)(E/E_1)^{-\Gamma} d\Gamma.
\end{equation}
Using the observed distributions $h(\Gamma)$ of Fig.\ref{fig:ScrGamdist} (right) we can numerically (and again non-parametrically as done for $J(S)$) calculate $H(E)$ and ${\cal E}(E)$. This result is shown by the solid red line in Fig.\ref{fig:EGB} (right).
We can also obtain an analytic expression for ${\cal E}(E)$ assuming
a Gaussian distribution for $\Gamma$ with mean value of $\Gamma_0$ and
dispersion $\sigma$, which gives
\begin{equation}\label{HofE2}
H(E)=\frac{1}{\sqrt {2 \pi} \sigma}\frac{e^{-E/E_c}}{E_1^2}\int_{-\infty}^\infty
e^{-(\Gamma-\Gamma_0)^2/2\sigma^2}(E/E_1)^{-\Gamma} g(\Gamma)d\Gamma.
\end{equation}
In any case, here we are interested primarily on the energy dependence part, which with $t\equiv(\Gamma-\Gamma_0)/\sqrt{2}\sigma$ can be written as

\begin{equation}\label{HofE3}
H(E)=\frac{1}{\sqrt \pi}\frac{e^{-E/E_c}}{E_1^2}(E/E_1)^{-\Gamma_0}\int_{-\infty}^\infty g(\Gamma_0+\sqrt{2}\sigma t)e^{-t^2-\sqrt{2}t\ln(E/E_1)^\sigma}dt.
\end{equation}

Defining $t_0=(\sigma/\sqrt{2})\ln (E/E_1)$ and setting $x=t+t_0$ we can rewrite the above equation as another Gaussian

\begin{equation}\label{HofE4}
H(E)=\frac{1}{\sqrt \pi}\frac{e^{t_0^2-E/E_c}}{E_1^2}(E/E_1)^{-\Gamma_0}\int_{-\infty}^\infty
g(\Gamma_0+\sqrt{2}\sigma (x-t_0))e^{-x^2}dx.
\end{equation}

Since $g(\Gamma)$ is a slowly varying function compared to the Gaussian we can expand it in Taylor series around $x=0$ or $\Gamma^\prime(E)=\Gamma_0 - \sigma^2 \ln (E/E_1)$ as $g(x)=g(0)+xg'(0)+ g''(0)x^2/2)$. The integrals of these three terms are $\sqrt{\pi}g(0), 0, \sqrt{\pi}g''(0)/4$. respectively.

\begin{equation}\label{HofE5}
H(E)=[g(\Gamma^\prime)+g^{\prime\prime}(\Gamma^\prime)/4](E/E_1)^{-\Gamma_0}\frac{e^{\epsilon(E)-E/E_c}}{E_1^2}\,\,\,\, {\rm with} \,\,\,\,
\epsilon(E)={\sigma^2 [\ln(E/E_1)]^2 \over 2},
\end{equation}
where $E_1=100$ MeV and $\Gamma^\prime(E)=\Gamma_0 + \sigma^2 \ln (E/E_1)$..
 With a little algebra we can write

 \begin{equation}
 g(x)=\sqrt{2}\sigma(x+a)/[1-e^{-b(x+a)}]\,\,\,\, {\rm with}\,\,\,\, a=(\Gamma_0-2)/(\sqrt{2}\sigma)-t_0 \,\,\,\, {\rm and} \,\,\,\, b=9.8\sigma.
 \end{equation}
We thus get $g(\Gamma^\prime)=\sqrt{2}\sigma\times a/(1-e^{-ba})$ and the second derivative
\begin{equation}
\label{d2gdx}
g^{\prime\prime}(\Gamma^\prime)=-\sqrt{2}\sigma\times {be^{-ab}\over (1-e^{-ab})^3}[2(1-e^{-ab})-ba(1+e^{-ab})].
\end{equation}
With $\sigma=0.28$ and $\Gamma_0=2.18$ we obtain $\sqrt{2}\sigma=0.4, b=2.73$ and $a=0.45-0.20\ln(E/E_1)$, which can be used in Eq. (\ref{HofE5}) to obtain the spectrum of EGB. The dashed blue line on the right Figure \ref{fig:EGB} gives the result from this analytic equation which is in excellent agreement with the numerical evaluation of the EGB spectrum.

\section{summary and discussion}
\label{summary}

We have used the rigorous non-parameteric methods of Efron-Petrosian and Lynden-bell (EP-L) to calculate the  gamma-ray luminosity function of FSRQs, and its cosmological evolution, directly from the  data on gamma-ray energy fluxes and spectral indexes, from the Fermi-LAT eight-year AGN catalog with $TS \geq 25$ and Galactic latitude $|b|>10^o$, and redshifts and optical magnitudes from Gaia counterparts 
of sources. This task requires data set with well defined gamma-ray and optical observational selection criteria and a joint determination of the optical and gamma-ray luminosity distribution and is evolution represented by the tri-variate luminosity function, $\Psi(L_\gamma,L_{\rm opt}, z)$, with proper accounting for the biases arising from the various data truncation produced by the observations. With our procedures we address first the selection criteria which allows us to use the EP-L method to  determine (1) the $L-z$ correlations, i.e.~the luminosity evolution in both wavelengths; (2) the intrinsic $L_\gamma-L_{\rm opt}$ correlation; (3) the forms of the luminosity function and the cosmological evolution of the co-moving density of FSRQs. Below we give a brief summary of the procedures and results for all blazars. 

\begin{enumerate}
    \item
{\it Flux-Index Relations:}

As is well documented by the Fermi collaboration, the gamma-ray flux threshold varies with the spectral index $\Gamma$, which truncates the data and induces a false correlation between the gamma-ray energy flux $S_{25}$ and $\Gamma$. Using the EP method we correct for this bias by determining the true correlation and determine a {\it correlation corrected} energy flux  $S^{\rm cr}_{25}$, its cumulative and differential distributions, as well as the unbiased distribution of $\Gamma$. This leads to a well defined threshold  essential for our goal. We also use these results to estimate the contribution of Blazars to the cosmic or extragalactic gamma-ray background radiation (EGB).

Spectral indexes in the optical range are also required for obtaining K-corrected optical luminosities. Gaia provides sufficient data to determine the optical indexes. We find no significant correlation between the optical flux and index and determine a well defined optical threshold for all blazars, and for FSRQs and BL-lacs separately.

\item

{\it Sample Selection and Completion}

Our goal is to  obtain the evolution of the luminosity function of the Blazars from the above observed sample, which is affected, and thereby truncated due to several observational selection effects. This tasks requires first and foremost "{\it complete}" samples, which means samples with well known  gamma-ray and optical flux limits. The sample of blazars in 4LAC  and those in GAIA have each well defined respective flux limits, but the joint sample is not complete because a small fraction of 4LAC sources have GAIA redshifts. This fraction is small for BCUs and BL Lacs, but is high (0.91) for FSRQs. The K-S test also indicates that the probability that the samples with redshift are drawn from the complete ``parent samples" is high only for FSRQs ($p$-value=0.81). Thus, we limit our analysis to FSRQs.

\item

{\it Luminosities and redshifts:}

From this data we obtain the three dimensional data ($L_\gamma, L_{\rm opt}, z$), which can be used for a complete determination of $\Psi(L_\gamma,L_{\rm opt}, z)$, including the correlations between all three variables (i.e.~the two luminosity evolution and the $L_\gamma-L_{\rm opt}$ correlation) as demonstrated previously for optical and radio samples of quasars (\cite{2011ApJ...743..104S,2013ApJ...764...43S} and \cite{2015IAUS..313..333P}). However, since here we are mainly interested in the gamma-ray characteristics we use the following simpler more accurate method.

\item
 {\it Optical and gamma-ray limited sub-samples:}

Instead of dealing with the tri-variate situation we separate the total sample with two flux thresholds into two distinct sub-samples one with gamma-ray only and one with optical only threshold, which we call gamma-ray and optically limited samples, respectively. This is done by evaluation of the two maximum redshifts that each source must have to be included in the sample. The sources with smaller maximum gamma-ray redshift go into the gamma-ray limited sample and vice versa. This allows separate determination of the LFs and their evolution as two distinct bi-variate functions, $\Psi (L_\gamma, z)$ and $\Psi(L_{\rm opt}, z)$.

\item

{\it Luminosity Evolutions:}

The above results show strong correlations between the two luminosities and redshift some of which is  due to the effects of the flux limits. Using the EP method we correct for this selection bias and find a significant intrinsic correlation approximated as $L \propto (1+z)^{k_{\gamma}}$ and  as in more realistic form in Eq. (\ref{gofz}), with 
$k_{\gamma}=5.1(5.8)$ for FSRQs.
These evolution indexes are similar to those found for radio luminosity evolution of quasars.

\item

{\it Luminosity-Luminosity Correlation:}

We also address the correlation between the gamma-ray and optical luminosites. As expected the observed luminosities show strong correlation because of their identical dependence on the luminosity distance, and the similar strong luminosity evolutions. We correct for the latter effect using the above evolution indexes, which removes some of the correlation but with a significant residual correlation due to the former effect; as determined by the large Pearson partial correlation coefficient (PPCC). To account for this effect and determine the intrinsic correlation we use the
technique introduced in \cite{2015IAUS..313..333P} to obtain the power-law index of the $L-L$ correlation, $\beta=0.68 \pm 0.04$ for FSRQs.
These is stronger correlations (with higher indexes) than what was found for radio$-$optical \cite{2015IAUS..313..333P} and mid-infrared$-$optical \citet{2019ApJ...877...63S} luminosities.

\item
{\it Luminosity Function and Density Evolution:}

Having defined a local luminosity uncorrelated with redshifts, we are then able to determine the local LF of FSRQs, $\psi(L_0)$, and the co-moving density evolution, $\rho(z)$. We find local gamma-ray LF that can be described by a smoothly broken  power law form, with power-law indexes somewhat different than those obtained by forward fitting methods (\cite[e.g.][]{2012ApJ...751..108A,2014ApJ...780...73A,2015ApJ...800L..27A}) and the the true redshift distribution, i.e.~the co-moving density evolution, which have plateau between redshift 1 and 2.5. 

\item

{\it Blazars and EGB:}

 Using the non-parametric determination of the distributions of the energy flux $S^{\rm cr}_{25}$ and spectral index $\Gamma$ we calculate directly, and  without resorting to redshift measurements or LF evaluation, the contribution of all Fermi-LAT blazars to the cosmic or extragalactic diffuse background gamma-ray intensity and spectrum. We find a spectral shape fairly similar to observation. However blazars up to flux threshold of Fermi-LAT $S^{\rm cr}_{25}=1.26\times 10^{-12}$ erg cm$^{-2}$ s$^{-1}$ can account for 32\%  of the observed EGB photons requiring extension of the flux to one or two  order of magnitude below the above threshold. However, such extension increases $J(S)$ without affecting the spectrum, which as evident from Figure \ref{fig:EGB} will raise the high energy end well above the observations. Agreement with the observed spectrum can be achieved if $E_c< 100$ GeV or if the lower flux sources have steeper spectra. A more likely possibility is that non-blazar sources, e.g. star forming galaxies, account for the rest of the EGB at low energies.

\end{enumerate}

The result from the above analysis can be used to investigate the origin and site of gamma-ray emission of the AGNs and its relation to the origin of emissions at other wavelengths. Such investigations are beyond the scope of this paper, but as an example we note that the similar and stronger luminosity evolution, and stronger intrinsic $L_\gamma-L_{\rm rad}$ correlation, at gamma-ray and radio regime as compared to optical evolution, and $L_\gamma-L_{\rm opt}$ correlation, may indicate a common origin of radio and gamma-ray emission (presumably from jets) as distinct from accretion disk emission of the optical emission.

\acknowledgments

We thank the anonymous referee for very helpful suggestions that help to improve the manuscript significantly. This work is partially supported by the National Natural
Science Foundation of China (NSFC-11703094 and NSFC-11863007), the National Key RD Program of China  2018YFA0404203.

\bibliographystyle{aasjournal}

\begin{longrotatetable}
\begin{deluxetable*}{lllll|llllllll}
\tablecaption{Sample for 1124 blazars. \label{sample}}
\tabletypesize{\scriptsize}
\tablehead{
\colhead{Source Name} & \colhead{Redshift} &
\colhead{$S_{25}$} & \colhead{$\Gamma$} &
\colhead{CLASS} & \colhead{Gaia Counterpart} &
\colhead{$G$} & \colhead{$G_{BP}$} &
\colhead{$G_{RP}$} & \colhead{$f_{\nu}$} & \colhead{$f_{\nu_{BP}}$}& \colhead{$f_{\nu_{RP}}$} & \colhead{$\alpha_{opt}$}\\
\colhead{} & \colhead{} & \colhead{erg cm$^{-2}$ s$^{-1}$} & \colhead{} &
\colhead{} & \colhead{} & \multicolumn{3}{|c|}{mag} & \multicolumn{3}{|c|}{mJy} &\colhead{}
}
\startdata
4FGL J0001.5+2113 & 1.106 & 1.926e-11 & 2.680 & fsrq & Gaia DR2: 2846735969108432384 & 18.484 & 18.597 & 17.800 & 0.133 & 0.129 & 0.199 & 1.023 \\
4FGL J0003.9-1149 & 1.310 & 3.211e-12 & 2.128 & bll & Gaia DR2: 2421598528234887680 & 19.322 & 19.661 & 18.617 & 0.062 & 0.048 & 0.094 & 1.556 \\
4FGL J0004.0+0840 & 2.057 & 2.487e-12 & 2.102 & bcu & Gaia DR2: 2746890967457065856 & 19.762 & 19.980 & 19.084 & 0.041 & 0.036 & 0.061 & 1.236 \\
4FGL J0004.4-4737 & 0.880 & 6.513e-12 & 2.415 & fsrq & Gaia DR2: 4978016560925910400 & 19.848 & 19.999 & 19.226 & 0.038 & 0.035 & 0.053 & 0.970 \\
4FGL J0005.9+3824 & 0.229 & 8.359e-12 & 2.667 & fsrq & Gaia DR2: 2880735415555982976 & 18.399 & 18.877 & 17.160 & 0.144 & 0.099 & 0.359 & 3.012 \\
4FGL J0006.3-0620 & 0.347 & 1.425e-12 & 2.171 & bll & Gaia DR2: 2442050857556048000 & 17.154 & 17.646 & 16.276 & 0.453 & 0.309 & 0.809 & 2.260 \\
4FGL J0006.4+0135 & 0.787 & 1.969e-12 & 2.095 & bcu & Gaia DR2: 2738768737624127360 & 19.612 & 19.700 & 18.636 & 0.047 & 0.047 & 0.092 & 1.600 \\
4FGL J0008.0+4711 & 0.280 & 1.827e-11 & 2.053 & bll & Gaia DR2: 393015758425406848 & 19.009 & 19.838 & 18.104 & 0.082 & 0.041 & 0.150 & 3.048 \\
\enddata
\end{deluxetable*}
\end{longrotatetable}


\end{document}